%% file: paper.tex
\documentclass[acmsmall,screen]{acmart}

\setcopyright{none}

\usepackage{xspace}
\usepackage{float}

\usepackage[utf8]{inputenc}

\usepackage{listings}
  \newcommand\maybecolor[1]{\color{#1}}
  \definecolor{dkblue}{rgb}{0,0.1,0.5}
  \definecolor{dkred}{rgb}{0.6,0,0}
  \include{lstfstar}
\usepackage{xparse}
  \NewDocumentCommand{\li}{v}{\lstinline{#1}}

\usepackage{xcolor,soul}
  \colorlet{hl}{gray!20} 
  \sethlcolor{hl}

\usepackage{tikz}
  \input{tango-colors.tex}
  \usetikzlibrary{calc}
  \usetikzlibrary{tikzmark}
  \usetikzlibrary{shapes.geometric}
  \usetikzlibrary{shapes,arrows}
  \usetikzlibrary{fit}          %
  \usetikzlibrary{positioning}
  \usetikzlibrary{backgrounds}  %
  \usetikzlibrary{shapes.callouts,decorations.text}

\newcommand\fstar{F$^\star$\xspace}
\newcommand\haclstar{HACL$^\star$\xspace}
\newcommand\lowstar{Low$^\star$\xspace}
\newcommand{\krml}{KaRaMeL\xspace}

\newcommand\noisestar{Noise$^\star$\xspace}
\newcommand\steel{Steel\xspace}

\newcommand{\sref}[1]{\S\ref{sec:#1}}
\newcommand{\fref}[1]{Figure~\ref{fig:#1}}
\newcommand{\tref}[1]{Table~\ref{table:#1}}
\newcommand{\lref}[1]{Listing~\ref{lst:#1}}

\newcommand{\comments}{false}
\usepackage{ifthen}
\ifthenelse{\equal{\comments}{true}}
{
  \DeclareRobustCommand{\son}[1]{ {\begingroup\color{red!60!black}{(Son) #1}\endgroup} }
  \DeclareRobustCommand{\af}[1]{ {\begingroup\color{green!60!black}{(Aymeric) #1}\endgroup} }
  \DeclareRobustCommand{\jonathan}[1]{ {\begingroup\color{blue!60!black}{(Jonathan) #1}\endgroup} }
}
{
  \DeclareRobustCommand{\son}[1]{ {} }
  \DeclareRobustCommand{\af}[1]{ {} }
  \DeclareRobustCommand{\jonathan}[1]{ {} }
}

\usepackage[frozencache]{minted}
  \setminted{fontsize=\footnotesize}

\newcommand\mytitle{Modularity, Code Specialization, and
Zero-Cost Abstractions for Program Verification}
\title[\mytitle]{\mytitle}

\begin{document}

\author{Son Ho}
\affiliation{\institution{Inria}\country{France}}
\email{son.ho@inria.fr}

\author{Aymeric Fromherz}
\affiliation{\institution{Inria}\country{France}}
\email{aymeric.fromherz@inria.fr}

\author{Jonathan Protzenko}
\affiliation{\institution{Microsoft Research}\country{USA}}
\email{protz@microsoft.com}

\begin{abstract}
For all the successes in verifying low-level, efficient, security-critical code, little has been said or studied about the structure, architecture and engineering of such large-scale proof developments.

We present the design, implementation and evaluation of a set of language-based techniques that allow the programmer to modularly write and verify code at a high level of abstraction, while retaining control over the compilation process and producing high-quality, zero-overhead, low-level code suitable for integration into mainstream software.

We implement our techniques within the F* proof assistant, and specifically its shallowly-embedded Low* toolchain that compiles to C. Through our evaluation, we establish that our techniques were critical in scaling the popular HACL* library past 100,000 lines of verified source code, and brought about significant gains in proof engineer productivity.

The exposition of our methodology converges on one final, novel case study: the streaming API, a finicky API that has historically caused many bugs in high-profile software. Using our approach, we manage to capture the streaming semantics in a generic way, and apply it ``for free'' to over a dozen use-cases. Six of those have made it into the reference implementation of the Python programming language, replacing the previous CVE-ridden code.
\end{abstract}

\maketitle

\section{Introduction}

Within the span of a few years, formal verification has gone mainstream. Previously
confined to academic circles, the idea of proving properties about security-critical code
is now widely accepted. Case in point: a major cloud company like Amazon will
pay for a full sponsored article in the Wall Street Journal, touting
the benefits of formal verification for its cloud computing unit~\cite{awswsj}.

Such security-critical code often lies on the critical path of larger subsystems;
users therefore expect security-critical code to be not only secure and reliable, but also fast.
To that effect,
programmers continue to resort
to low-level programming idioms and manual memory management, which allows
them to exert fine-grained control on the structure of their code, and hence
squeeze every last inch of performance out of it~\cite{warren2013hacker},
sometimes directly leveraging hardware
facilities to do so~\cite{intelaesni,fast-curve25519}.
This unfortunately comes at a cost; taming the complexity of such programs
is error-prone, leading to abundant mistakes with dire
consequences~\cite{CVE-2017-5715,CVE-2017-5753,CVE-2014-0160,polybug,polybug2,firefox-bug,mouha2018finding,sha3-bug,jason-found-a-bug,bitcoin-merkle-bug,aes-gcm-bug,openssl-poly1305-bug1,openssl-poly1305-bug2,openssl-poly1305-bug3,springboard}.

Aiming to address this problem, formal verification practitioners have thus
focused on code that is both security-critical and low-level.
Success stories include
verified cryptography,
with e.g., \haclstar/EverCrypt~\cite{hacl,haclxn,evercrypt}
(integrated into the Linux kernel, Mozilla Firefox, and the Tezos
blockchain), or Fiat Cryptography~\cite{fiat-crypto} (integrated into Google's
BoringSSL cryptographic library); verified parsers,
with e.g., EverParse~\cite{everparse,everparse3d}
(integrated into Microsoft's Hyper-V network virtualization stack);
verified kernels, such as
CertiKOS~\cite{Gu:CertiKOS:2015,Gu:2016:CEA:3026877.3026928}
or seL4~\cite{klein2009seL4}; and many more.

But for all the success stories, it remains a technical challenge to author and
verify a system that is simultaneously large-scale, low-level, and performant.
Large-scale verification projects abound; one must only think of e.g.,
Mathlib~\cite{mathlib2020},
a large collection of mathematical proofs and theorems written in Lean3, which
recently crossed the million-line threshold.
Large-scale \emph{and} low-level verified projects are not unheard of:
seL4~\cite{sewell2013translation},
based on a dialect of Haskell, or CertiKOS~\cite{Gu:2016:CEA:3026877.3026928}, written in Coq, both demonstrate
that one can write non-trivial pieces of software (such as OSes) that deal with
low-level concerns and deliver \emph{reasonable} performance.
But when it comes to large-scale, low-level \emph{and} competitively performant verification
projects, few candidates come to mind.
One reason is that verification remains onerous: expert proof engineers are
rare, and their task is hard enough; as such, advances in proof engineering and
reusable abstractions are badly needed to increase productivity.
Nowhere is this more salient than when trying to verify low-level code.
Verification calls for high-level abstractions and extreme
modularity, while low-level efficient code calls for breaking up those very
abstractions barriers. This, in our opinion, has hindered the development of
large-scale, low-level, and efficient libraries.

High-level abstractions are well-known to functional programmers; they may
include type-level abstraction and polymorphism; module interfaces and functors;
type classes. They are also known to the mythical ``real-world'' programmers:
templates and concepts in C++, or traits in Rust, also support modularity in the
large.
But sooner or later, the programmer will, on the quest to ultimate performance,
pull low-level tools from their arsenal. The infamous C preprocessor oftentimes
makes an appearance, with tricks so frightening that basic decency prevents us
from describing them here~\cite{cpp-tricks}.
And these are not just simple, straightforward patterns
such as loop unrolling; entire polymorphic data structures
are emulated using the C preprocessor. This route usually ends in pain and
suffering, with unmaintainable code, subtle mistakes, and generally, the
inability to reason about such code.
There is thus a tension between going low-level for efficiency, and introducing
high-level concepts, abstraction and modular boundaries to make the code easier to
reason about. This tension is heightened in the context of verification: the
need for modular, high-level code is even greater, so as to ease
verification; but the pressure for efficient, low-level code is also stronger, to
meet practitioners' performance requirements and thus give our verified code the
chance to be integrated and then deployed in mainstream software.

In this paper, we set out to have our cake and eat it. That is, to have efficient,
low-level, verified code, and to do so \emph{at a large scale},
in a verified software project that exceeds 100,000 lines of verified code.
Worded differently, we want to
reconcile the modularity of, say, SML or OCaml's functors, or Haskell's type
classes, with the efficiency of Rust traits and the infamous ``zero-cost abstraction'' of C++
templates, \emph{for verified code}. And we want to
have our cherry on top of the cake: we set out to do so \emph{without extending
the Trusted Computing Base (TCB)} of the tools we use.
We design, implement and evaluate our techniques within the \fstar
dependently-typed proof assistant, which culminate in the following
contributions.

First (\sref{encoding}), we propose a proof engineering methodology
that allows one to structure their verified code as they would, say, with functors,
all the while still producing idiomatic, low-level code with
readable functions and no runtime overhead of any kind.

Second (\sref{tactic}), we observe that using this methodology is burdensome in
practice, because structuring the code to fit our proof engineering pattern requires a
fair amount of bookkeeping. We thus automate our methodology by designing
a DSL that guides an automated code-rewriting transformation which automatically applies
the pattern from \sref{encoding} to the user's code.
In practice, this allows the user to write their code in a modular, high-level,
natural style that emulates ML's functors, while
relying on inlining and partial evaluation to eliminate the high-level
abstractions and make our discipline truly, a zero-cost abstraction.
The DSL is interpreted via meta-programming, specifically, via elaborator
reflection; in essence, we script the compiler, and add an early compilation
stage that takes our functor DSL and evaluates it away.
The techniques we introduce are implemented in user-space, meaning
we do not modify the compiler and leave the TCB intact, so as to provide the
same guarantees as code written without our libraries.

Third (\sref{case-studies}), we explain how several algorithms previously released via the \haclstar
and EverCrypt projects were, in reality, relying on our techniques to scale up,
and to avert engineering and usability disasters. We review a series of
case studies and show how several cryptographic primitives can be implemented
using our DSL so as to maximize code sharing and minimize maintenance.

Fourth (\sref{streaming}), and final, we examine a large case study: the streaming API, a
cryptographic construct that transforms an unsafe, block-based algorithm into
a safe, high-level API by means of an internal buffer. With our DSL, we write
a generic streaming API once, then instantiate it ``for free'' over any unsafe
block-based algorithm. Out of a dozen instantiations of our streaming
functor, six have been integrated into the reference implementation of the
Python programming language.
This case study is a contribution on its
own: to the best of our knowledge, no one had precisely described, captured with
dependent types and implemented generically what it means to turn a block-based
algorithm into a streaming API.

Our evaluation section quantifies the improvements in
programmer productivity and effectiveness stemming from the use of our methodology.
We have evaluated our techniques on the \haclstar project, and found that they
were the key ingredient that allowed \haclstar to cross the barrier of 100,000
lines of verified source \fstar code. Without our work, modularizing and scaling up the
codebase would have been impossible.

We conclude and observe that while our case studies focus on cryptographic
code, our techniques are general and can be applied to data structures, or more
generally, any situation that calls for modular proofs of low-level programs, as
evidenced by our choice of running example (\sref{encoding}).

\section{Background}
\label{sec:background}

In this section, we introduce the background required to understand our methodology.
We start our presentation by an overview of our verification environment: the \fstar proof
assistant (\sref{fstar}). We then present a well-known technique to encode
functors with dependent types (\sref{depTypesFunc}) and that we build upon
in the later sections.

\subsection{\fstar, \lowstar, Meta-\fstar}
\label{sec:fstar}

\textbf{\fstar} is a state-of-the-art verification-oriented programming language. Hailing
from the tradition of ML~\cite{ml}, \fstar features dependent types, refinement types, and a
user-extensible effect system~\cite{indexedeffects},
which allows reasoning about IO, concurrency,
divergence, various flavors of mutability, or any combination thereof. For
verification, \fstar uses a weakest precondition calculus based on Dijkstra
Monads~\cite{mumon,dm4free}, which synthesizes verification conditions that are then
discharged to the Z3 SMT solver~\cite{moura2008z3}. Proofs in \fstar typically
are a mixture of manual reasoning (calls to lemmas), semi-automated reasoning
(via tactics~\cite{fstar-meta}) and fully automated reasoning (via SMT).

\textbf{\lowstar} is a subset of \fstar that exposes a carefully curated subset
of the C language.  Using \fstar's effect system, \lowstar models the
C stack and heap, and allocations in those regions of the memory. \lowstar
also models data-oriented features of C, such as arrays, pointer arithmetic,
machine integers with modulo semantics, \li+const+ pointers, and many others via
a set of distinguished libraries. Programming in \lowstar guarantees spatial
safety (no out-of-bounds accesses), temporal safety (no double frees, no
use-after free) and a form of side-channel resistance~\cite{hacl,lowstar}. All
of these guarantees are enforced statically and incur no run-time checks.
To provide a flavor of programming in \lowstar, we present the
\li+swap+ function below.
We first focus on the various typical \lowstar features of
this function signature.

\begin{minted}{fstar}
let swap (x y : pointer U32.t) : ST unit (requires ...) (ensures ...) =
  let xv = deref x in let yv = deref y in upd x yv; upd y xv
\end{minted}

Functions in \lowstar are annotated with their return effect, in this case
\li+ST+, which indicates that the function may perform heap allocations%
\footnote{\lowstar actually distinguishes two stateful effects, \li+Stack+ for functions
which only allocate on the stack (no memory leaks), and \li+ST+ for functions which also
allocate on the heap. In this paper, we only use \li+ST+
for the purpose of simplicity.}.
Functions without a return effect are understood to be total.
The input parameters have type \li+pointer U32.t+, i.e., pointers to
32-bit unsigned machine integers with modulo
semantics. Functions are specified using pre and post-conditions, which we omit
here and whose
explanation we defer until \sref{streaming}.
Finally, the implementation of \li+swap+ simply dereferences
\li+x+ and \li+y+ (\li+deref+), then updates them while swapping their
values (\li+upd+).

\textbf{Erasure and extraction in \fstar} follows Letouzey's extraction
principles for Coq~\cite{letouzey2002new}. After type-checking and performing
partial evaluation, \fstar \emph{erases} computationally-irrelevant code and
performs extraction to an intermediary representation dubbed the ``ML AST''.

For erasure, \fstar eliminates type refinements, pre- and post-conditions, and
generally replaces computationally irrelevant terms with units.
\fstar also
removes calls to (pure) unit-returning functions, which means that calls to lemmas are
also eliminated.
For extraction, \fstar ensures that the ``ML AST'' features only prenex
polymorphism (i.e., type schemes), and that it is annotated with classic ML
types.
In the context of this paper, we are only concerned with the generation of C
code, which is possible only on a subset of the ``ML AST''; when extracting for
C, a battery of checkers verifies that the code is in the proper subset.

\textbf{\krml}~\cite{lowstar} compiles the ``ML AST'' to \emph{readable},
\emph{auditable} C by using a series of small, composable passes. The \krml
preservation theorem~\cite{lowstar} states that the safety guarantees in
\lowstar carry over to the generated C code.
We show below the result of compiling \li+swap+ to C.

\begin{minted}{c}
void swap(uint32_t *x, uint32_t *y) { uint32_t xv = *x; uint32_t yv = *y; *x = yv; *y = xv; }
\end{minted}

\subsection{Encoding Functors With Dependent Types}
\label{sec:depTypesFunc}

We now illustrate the challenge of combining generic, modular programming (good
for proofs) with low-level compilation (good for efficiency). We start with a running
example that we will reuse in \sref{encoding}: an imperative key-value map implemented using an
associative list.
For simplicity of exposition, we use standard algebraic
datatypes, such as \li+list+. \lowstar features low-level data structures,
notably linked lists; however, these would
significantly complicate our running example with notions of memory footprints
and memory reasoning. We thus stick with \li+list+ for the
paper, and provide a complete low-level example relying on linked lists in the supplementary material.

To enable code reuse, we wish to make the associative list generic in the type of its keys
and values.
If we were to use a language like OCaml or Haskell we would naturally
implement this map by using a functor or typeclasses.
\lref{map_functor} illustrates this with an OCaml functor named \li+MkMap+,
which takes an
argument \li+EqType+ containing a
type for keys \li+k+, and a corresponding decidable equality. The \li+MkMap+
functor implements
\li+find+ using a loop and mutable references, generically, for any type of keys
\li+k+ and corresponding equality \li+eq+.
We want to attain the same modularity when verifying code in a
prover like \fstar. As a first attempt, we can reuse a well-known
technique~\cite{rossberg2014f,macqueen1986using} to encode this OCaml functor using
dependent types (\lref{map_functor}, right).
The \li+Map+ module signature becomes a record \li+map+, and the type
\li+k+ of keys becomes a record field. Since this is a dependent record,
\li+eq+ may refer to \li+k+.
We implement the \li+MkMap+ functor with the \li+mk_map+ function, which
receives an instance of \li+eq_type+ along with a type \li+a+.
The return type of \li+mk_map+ uses a refinement: a value \li+m:map+ has type
\li+m:map a{m.k == e.t}+ if it satisfies the logical predicate \li+m.k == e.t+;
this equation exactly encodes the condition \li+type key = E.t+ of the OCaml
code (line~\ref{MkMap:eq}).
Finally, \lowstar uses a special \li+while+ combinator for loops, which takes
two closures as inputs, for the loop condition and the loop body respectively;
the implementation of \li+find+ otherwise mimics its OCaml counterpart.
As the code is stateful, i.e., it lives in the \li+ST+ effect, it requires
annotations such as pre- and post-conditions; at this point in the paper we are
concerned with the shape of the code and not its correctness, we thus omit them
for the purpose of simplicity.

\begin{figure}
\begin{minipage}[t]{0.48\textwidth}
\centering
\begin{minted}[linenos,mathescape=true,escapeinside=~~,texcomments]{ocaml}
module type Map = sig
  type k
  val find: k -> (k * 'a) list -> 'a option
end

module type EqType = sig
  type t
  val eq: t -> t -> bool end

module MkMap (E : EqType) :
  Map with type key = E.t = struct~\label{MkMap:eq}~
  type k = E.t
  let find x ls =
    let b = ref true in
    let lsp = ref ls in
    while !b do
      match !lsp with
      | [] -> b := false
      | (x', _) :: tl ->
        if E.eq x x' then b := false
        else lsp := tl done;
    match !lsp with
    | [] -> None | (_, y) :: _ -> Some y
end
\end{minted}
\end{minipage}
\begin{minipage}[t]{0.46\textwidth}
\centering
\begin{minted}[linenos,mathescape=true,escapeinside=~~]{fstar}
type map (a : Type) = {
  k: Type;
  find: k -> list (k * a) -> ST (option a) ... }

type eq_type = {
  t: Type;
  eq: t -> t -> bool; }

let mk_map (e : eq_type) (a : Type) :
  m:map a{m.k == e.t} = {
  k = e.t;
  find = (fun x ls ->
    let b = alloc true in
    let lsp = alloc ls in
    while (fun () -> !* b)
      (fun () ->
        let ls = !* lsp in
        match ls with
        | [] -> upd b false
        | (x', _) :: tl ->
          if e.eq x x' then upd b false
          else upd lsp tl);
    match !* lsp with
    | [] -> None | (_, y) :: _ -> Some y) }
\end{minted}
\end{minipage}
\captionof{listing}{An associative map implemented in OCaml (left) and \fstar (right)}
\label{lst:map_functor}
\end{figure}

Even when assuming that all data structures are suitably low-level,
the issue remains that the implementation
of \li+find+ manipulates dictionaries (e.g., instances of \li+eq_type+).
Note that this is not specific to our encoding: we would have the same problems
had we used functors or typeclasses, and this is the case for the OCaml
implementation of the map.
Dictionary-passing is problematic because it has a cost at runtime.
Worse, our implementation doesn't fit in the \lowstar subset and thus can't be
extracted to C; indeed, the resulting code would manipulate records with fields
containing types, which is not supported in C. We show in the next section how we solved
this problem.

\section{Writing Low-Level, Modular Code}
\label{sec:encoding}

We showed in \sref{depTypesFunc} how one can achieve the same level of
modularity and genericity in \fstar as in a regular, high-level programming language like
OCaml, by encoding functors with dependent types by means of an already known technique.
But now we ask: how can one turn this into idiomatic, efficient low-level code?
In the coming section, we answer by introducing new methods which build
upon the technique explained in \sref{depTypesFunc}.
We stick to the same running example, that is an imperative key-value map, and for the
purpose of illustration, we assume once again that all data structures, such as
\li+list+, are suitably low-level.

\subsection{Making Functors Zero-Cost: A First Attempt}
\label{sec:zerocost}

We now present a first naive technique that allows the user to successfully
generate specialized \lowstar code (i.e., without dictionary-passing), at the
expense of code size explosion.
The key idea is to perform partial evaluation at extraction time to inline all
uses of \li+eq_type+ (and \li+a+).
To do so, we can leverage the \fstar \emph{normalizer} to symbolically reduce terms. 
The normalizer is not an \fstar specificity; it is at the core of dependent
type systems, and therefore a mandatory component of the type-checker
of a dependently typed language. As such, this component is part of the
TCB of type-theory-based proof assistants.

The user proceeds as follows. First, they pick concrete values for the functor
arguments. In our example
(Listing~\ref{lst:find:inst}), the user picks \li+str_eqty+ and \li+int+ for the
\li`mk_map` parameters \li+e:eq_type+ and \li+a:Type+, respectively.
Then, the user applies those
arguments to the functor itself, hence defining an \textbf{\emph{i}}nstantiated version
of \li+find+, dubbed \li+ifind+ (line~\ref{lst:find:inst:mk_map}).
The normalizer then kicks in and $\beta$-reduces the application of \li+ifind+
to its concrete arguments. By inlining the body of
\li+find+, then by simplifying some terms like the projection
\li+str_eqty.eq+,
all uses of records inside \li`ifind`
are removed; the resulting specialized \li`ifind` is indistinguishable from a
direct, monomorphic implementation of \li`find`. We show the result of partial
evaluation in \lref{find:inst}.

\begin{figure}
  \centering
\begin{minted}[linenos,mathescape=true,escapeinside=~~,texcomments]{fstar}
(* Map instantiation *)
let str_eqty : eq_type = { t = string; eq = String.eq; }
let ifind = (mk_map str_eqty int).find ~\label{lst:find:inst:mk_map}~

(* After partial evaluation *)
let ifind (x: string) (ls: list (string * int)): option int =
  let b = alloc true in let lsp = alloc ls in
  while (fun () -> !* b)
    (fun () -> let ls = !* lsp in match ls with
      | [] -> upd b false | (x', _) :: tl -> if String.eq x x' then upd b false else upd lsp tl);
  match !* lsp with | [] -> None | (_, y) :: _ -> Some y
\end{minted}
\captionof{listing}{\li+find+ after instantiation (top), then partial evaluation
(bottom)}
\label{lst:find:inst}
\end{figure}

This approach therefore allows us to turn our functors into zero-cost abstractions.
The caveat with this style, however, is that
\emph{every single function} needs to be inlined, except for the top-level
functions that make up the API entry points. This is fine for our small example;
but in a real-world development, this leads to both code size explosion (we insert a
copy of \li+find+'s body at each call-site), and unacceptable code quality
(implementing an algorithm as a single 20,000-line C function is generally frowned upon).
To illustrate this more concretely, we introduce in~\lref{device} a client of \li+find+, known as
a ``device'', a high-level data structure used in communication protocols to
store a map from peer identifiers to session keys, i.e., a map from unique
participant identifiers to the cryptographic keys used for secure communications.

\begin{figure}
\begin{minipage}[t]{1\textwidth}
\centering
\begin{minted}[linenos,mathescape=true,escapeinside=~~,texcomments]{fstar}
type dv = {
  pid : Type;
  send : pid -> list (pid * ckey) -> bytes -> option bytes;
  recv : pid -> list (pid * ckey) -> bytes -> option bytes; }

type cipher = { enc : ckey -> bytes -> bytes; dec : ckey -> bytes -> option bytes; }
let mk_dv (m : map ckey) (c : cipher) : d:dv{d.pid == m.k} = {
  pid = m.k;
  send = (fun id dv plain  -> match m.find id dv with | None -> None | Some sk -> Some (c.enc sk plain));
  recv = (fun id dv secret -> match m.find id dv with | None -> None | Some sk -> c.dec sk secret) }
\end{minted}
\end{minipage}
\captionof{listing}{Implementation in \fstar of a peer device for a secure
  channel protocol}
\label{lst:device}
\end{figure}

A device should implement two functions to communicate with participants.
The \li+send+ function takes as arguments a peer identifier \li+id+, a map from peer identifiers
to cryptographic keys (of type \li+ckey+), and a message \li+plain+.
It looks up the key associated to \li+id+, and finally uses it to encrypt \li+plain+.
The \li+recv+ function performs the dual operation, i.e., it searches for the key to
decrypt a message received from a known peer.
The choice of the peer identifier type \li`pid` is orthogonal to the implementation
of a device \li`dv`; we can therefore write a generic implementation parametric
in \li`pid` and accordingly in the map from peer identifiers to cryptographic keys.
Furthermore, this device can be useful in a variety of contexts and with a range 
of ciphersuites, and should thus be independent of the specifics of any cryptographic
encryption algorithm:
we also parameterize the implementation \li+mk_dv+ with the encryption
and decryption functions, encapsulated in a record of type \li+cipher+. 

Equipped with a generic device, we can, as in the map example, 
instantiate it for a specific choice of peer identifiers and cryptographic functions,
before applying partial evaluation to get specialized code which does not
manipulate dictionaries.
Unfortunately, doing so would lead us into the pitfall we mentioned earlier,
where the code for \li+find+ is duplicated in both the instantiations for
\li+send+ and \li+recv+.
In our experience interacting with maintainers of some of the most popular
open-source projects, such aesthetic \emph{faux-pas} are bad enough that
a practitioner will dismiss our code as `not serious' and `too verbose',
raising barriers to its integration to existing codebases.
In this regard, we insist on the fact that the \haclstar code, part of which we applied
our methodology on (\sref{case-studies}), was deployed in real-world projects such
as the NSS library or Python.

\subsection{A General Rewriting Pattern for Fine-Tuned Code Generation}
\label{sec:no_functor}

When specializing functions like \li+send+ and \li+recv+, we want them to call the
same \emph{specialized version} of \li+find+, rather than duplicate the body of
\li+find+.
In effect, we want to perform whole-program specialization (in the style of
MLton~\cite{weeks2006whole}) while preserving the shape of the \emph{static} call-graph (in order to
give the programmer enough control so as to generate palatable code).
To do so, we propose a \emph{modular} approach that allows us to rewrite
\emph{each function in isolation}
without knowing yet how the function
will later be instantiated, all the while avoiding the need for inlining
everything. We proceed as follows.
Instead of using a dependent record, for each function, we add additional parameters that
stand in for the callees that need to be specialized; and we re-implement the
function body to refer to those arguments. For instance, \li+send+ and
\li+recv+ become the \li+mk_send+ and \li+mk_recv+ functions in \lref{device:no_functors}.
Note that a function is parameterized with \emph{exactly} its callees: for
instance \li+send+ is parameterized by \li+enc+ but not \li+dec+,
while it is the converse for \li+recv+.
We intentionally refrain from using a record (``functor'')-based
encoding like in the previous sections: this would rapidly lead to a
proliferation of type definitions, as there would typically be one record per
definition. This would make both programming and maintaining our codebase tedious,
as the addition or modification of any element in the record would require changing
all occurences across the call-graph.

\begin{figure}
\begin{center}
\begin{minted}[linenos,mathescape=true,escapeinside=~~,texcomments]{fstar}
let mk_send (pid : Type)
  (find : pid -> list (pid * ckey) -> option ckey) (enc : ckey -> bytes -> bytes)
  (id : pid) (dv : list (pid * ckey)) (plain : bytes) : option bytes =
  match find id dv with | None -> None | Some sk -> Some (enc sk plain)

let mk_recv (pid : Type)
  (find : pid -> list (pid * ckey) -> option ckey) (dec : ckey -> bytes -> option bytes)
  (id : pid) (dv : list (pid * ckey)) (plain : bytes) : option bytes =
  match find id dv with | None -> None | Some sk -> dec sk plain
\end{minted}
\captionof{listing}{Parameterizing \li`send` and \li`recv` by their callees}
\label{lst:device:no_functors}
\end{center}
\end{figure}

Anticipating a bit on the automation we introduce in \sref{tactic},
we request that the polymorphism be prenex, i.e., that all type
parameters be captured by the first argument; this does not restrict
expressivity, and allows us to avoid having extra type-level dependencies across
function arguments which would be difficult to automatically handle.
More specifically, we keep the generic types in a record, that we call the
``index'' and make the first parameter of the function.
This index must capture all the choices of parametricity for the types.
In practice, as we'll see in concrete, real-world examples in \sref{case-studies},
we often pick the index to be an enumeration, but this is not a requirement of our approach.
The index can also range over an infinite number of elements, as is the case
for the generic type \li`pid` in \lref{device:no_functors}.
In this specific example, since we are only parametric in one type,
we dispense with a record type and parameterize our functions over \li+pid+
directly.

We also apply this approach to the \li`find` function previously presented.
This function is parametric in two types: the type of keys \li`k`,
and the type of values \li`v` of the map.
We collect both types in a record of type \li`mindex`, which becomes the first
argument of \li`mk_find`, presented in \lref{find:mk}.

\begin{figure}
\begin{center}
\begin{minted}[linenos,mathescape=true,escapeinside=~~,texcomments]{ocaml}
type mindex = { k : Type; v : Type }

let mk_find (i : mindex) (eq : i.k-> i.k -> bool) (x : i.k) (ls : list (i.k * i.v)) : option i.v =
  let b = alloc true in let lsp = alloc ls in
  while (fun () -> !* b)
    (fun () -> let ls = !* lsp in
     match ls with | [] -> upd b false
                   | (x', _) :: tl -> if eq x x' then upd b false else upd lsp tl);
  match !* lsp with | [] -> None | (_, y) :: _ -> Some y)
\end{minted}
\captionof{listing}{Rewriting \li+find+ to follow a systematic pattern}
\label{lst:find:mk}
\end{center}
\end{figure}

Importantly, we drop the ``functor'' encoding for the functions but not the
types, i.e., we use a record which holds all the type parameters.
Keeping this encoding for types doesn't lead to the same proliferation of
records as for functions. Indeed, type parameters tend to be fewer, change
less often, and their
parameterization tends to be more uniform accross functions.

We finally show an instantiation of those generic definitions in
\lref{device:inst_sol}, where \li+aes_enc+ and \li+aes_dec+ are
encryption/decryption functions for AES-GCM, one of the most widely
used authenticated encryption algorithm. We omit their implementation, which
is irrelevant for presentation purposes; they can be provided by a separate
cryptographic library.
With this new encoding, we can individually unfold the definitions of \li+mk_find+,
\li+mk_send+ and \li+mk_recv+ before simplifying the projections over record fields,
e.g., \li+i.k+, while preserving the call graph; we show the result of the partial
evaluation in \lref{device:inst_sol}. Note in particular that the definition of
\li`mk_find` is not inlined in the resulting \li`isend` and \li`irecv`;
both functions instead call the specialized \li`ifind`.
Our goals are met: we have described a general rewriting pattern that allows us
to write generic implementations, that can be specialized for a choice of types (the ``index''),
while preserving the shape of the static call-graph and hence produce
high-quality low-level code.

\begin{figure}
\begin{minipage}[t]{1\textwidth}
\begin{center}
\begin{minted}[mathescape=true,escapeinside=~~,texcomments]{fstar}
(* Instantiation *)
let i = { k = string; v = ckey; }
let ifind = mk_find i String.eq
let isend = mk_send string ifind aes_enc
let irecv = mk_recv string ifind aes_dec

(* After partial evaluation *)
let ifind x ls =
  let b = alloc true in let lsp = alloc ls in
  while (fun () -> !* b) (fun () -> let ls = !* lsp in match ls with
     | [] -> upd b false | (x', _) :: tl -> if String.eq x x' then upd b false else upd lsp tl);
  match !* lsp with | [] -> None | (_, y) :: _ -> Some y)

let isend = id dv plain -> match ifind id dv with | None -> None | Some sk -> Some (aes_enc sk plain)
let irecv = id dv secret -> match ifind id dv secret with | None -> None | Some sk -> aes_dec sk secret
\end{minted}
\end{center}
\end{minipage}
\captionof{listing}{Instantiation (top) and partial evaluation (bottom) of the map and device functions}
\label{lst:device:inst_sol}
\end{figure}

\paragraph{Discussion}
Previous work in \fstar used a precursor to the techniques we present in this
section. In particular, \haclstar~\cite{hacl,haclxn} made heavy use of specialization and
partial evaluation to factor out large pieces of code, for instance by writing a single
generic implementation of Poly1305 for three variants (C, C+AVX, C+AVX2).
However, this early style had two issues.
First, it
led to code size explosion due to excessive inlining, which was solved
by manually introducing alternating levels of generic and specialized functions,
a tedious and time-consuming task.
Second, it relied on closed enumerations (i.e., an inductive with constant
constructors) as opposed to the open-ended
``indices'' that we introduce in the present section.
The first point is addressed by our DSL, the rewriting tactic and the
systematic higher-order pattern it produces. Regarding the second point,
parameterizing over closed enumerations is a legacy style (\sref{case-studies})
that is acceptable
as long as the user is adamant that no further cases will be added. Indeed, adding a
new case to the enumeration entails a re-verification of the generic code,
affecting modularity. We strongly encourage users to try to define a generic
index type (i.e., at type \li+Type+), which provides more flexibility,
modularity, and allows the user to trivially add new specializations without
affecting the generic code. This requires, however, more thought on the part of
the user to correctly define the index type.

Closer to the present work, \noisestar~\cite{noisestar} uses a mix of the encodings
presented in sections \ref{sec:zerocost} and \ref{sec:no_functor} to make the
implementation generic in, say, the cryptographic primitive implementations or the peer
identifiers. More precisely, it uses the idea of
writing generic \li+mk_+ functions that are later specialized, as we do in this section,
but where the \li+mk_+ functions are parameterized in a style
closer to the ``functor'' parameters of \sref{zerocost} (i.e., without an
index), because it didn't leverage the automation that we introduce in the next section.
Importantly, the \noisestar paper does not detail nor claim this technique due to lack of
space, and rather focuses on the use of partial evaluation on code not written in a
``functor'' style.
As such, the present paper is for us an opportunity to detail in one place the culmination
of all the techniques which were introduced to make the Everest~\cite{everest-snapl}
project scale up to its current size.

\section{Static Call-Graph Rewriting with Meta-Programming}
\label{sec:tactic}

\newcommand\kw[1]{\ensuremath{\textsf{#1}}}
\newcommand\kfun{\lambda\;}
\newcommand\klet{\kw{let\;}}
\newcommand\kreach{\kw{reach}}
\newcommand\kspec{\kw{spec}}
\newcommand\kindex{\kw{index}}

In the previous section, we identified a programming pattern that allowed us
to modularly write verified code, in a way reminiscent of ML functors,
by rewriting our low-level functions into a higher-order form that lends itself
to code specialization via partial application.
In practice, manually writing code which uses this pattern requires a fair
amount of tedious, administrative work.
In this section, we thus set out to relieve the user from this burden by designing a small
DSL, to be more precise a subset of \lowstar extended with a mechanism of annotations,
by which the user can write code in a natural style before calling a rewriting
procedure which automatically turns this code into a higher-order form.
To do so, we i) propose a small usability tweak to make parameterization
easier, then ii) formally define our rewriting rules, and iii) devise a frontend
language that allows the user to express their intent via a mechanism of
annotations. Our rewriting rules are interpreted by a custom pre-processing
phase implemented via elaborator reflection, i.e., ``scripting the compiler''. In
effect, we are adding a user-defined early compilation stage.

\subsection{A Declarative Style for Callee Arguments}
\label{sec:assume:val}

The higher-order, rewritten functions presented in \sref{no_functor}
allow us
to write low-level, verified code in a modular fashion.
However, there remains a usability problem. The functions
that we parameterize over, like \li+eq+, need to be brought in scope frequently, as
\li+eq+ has many callers. This
is currently achieved by making \emph{every function} in our development that
needs it
parametric over \li+eq+, which incurs a non-trivial amount of
boilerplate. Even worse, in the case of an actual algorithm, e.g., Curve25519 (\sref{curve}),
we parameterize the algorithm over a dozen operations.
Asking the user to add as many arguments to every declaration
\emph{and} call-site would be too onerous.

To alleviate these concerns, we propose to adopt a more declarative style.
For presentation purposes, let us reuse the
map example from the previous section. 
Instead of explicitly parameterizing the definitions (e.g., \li`find`)
with their generic parameters (e.g., the decidable equality \li`eq`),
we introduce the parameters of our implementations as top-level declarations
annotated with the \li`assume` qualifier, as shown in \lref{find_dsl}. 
We achieve the same effect as before: the declaration is in scope for
our entire development. But this time, we avoid the syntactic
overhead.
Once this declaration is in the scope of \li`find`, it can be freely
used and referred to in the body of the function. In practice, the index is an
implicit argument, which further reduces the syntactic burden.

\begin{figure}
\begin{center}
\begin{minted}[linenos,mathescape=true,escapeinside=~~,texcomments]{fstar}
type mindex = { k : Type; v : Type }
assume val eq (i : mindex) : i.k -> i.k -> bool

let find (i : mindex) (x : i.k) (ls : list (i.k * i.v)) : option i.v =
  let b = alloc true in let lsp = alloc ls in
  while (fun () -> !* b)
    (fun () -> let ls = !* lsp in
     match ls with | [] -> upd b false
                   | (x', _) :: tl -> if eq i x x' then upd b false else upd lsp tl);
  match !* lsp with | [] -> None | (_, y) :: _ -> Some y
\end{minted}
\end{center}
\captionof{listing}{Hoisting callee arguments from \li`find`}
\label{lst:find_dsl}
\end{figure}

With this approach, changing the signature of \li`eq` becomes less dreary.
Instead of performing modifications in all functions relying on \li`eq`, it
suffices to tweak its top-level declaration. The reader might wonder
why one would need to change the type of \li`eq`; while this example is
overly simple for presentation purposes, making minor modifications to
specifications to, say, add a missing invariant or fix a mistake in a precondition
is common when doing incremental verification. Leveraging \fstar's SMT-backed
automation, small changes to the callee often do not require modifying the callers.

An assumed declaration in \fstar is tantamount to introducing a hole in our code.
Trying to generate C code containing such a hole would lead to C extern declarations,
and raise compilation failures unless an external definition is provided by the user.
In the following section, we will describe how to fill this hole, and ensure that
the provided definition matches the assumed function type.

\subsection{Static Call-Graph Rewriting}
\label{sec:tactic:rewriting}

While hoisting callee arguments to assumed top-level declarations reduces code clutter,
it only alleviates some of the burden that a programmer is facing when using our methodology.
Relying on top-level assumed declarations for callees is not always desirable. In our map and device
example, while \li`send` and \li`recv` are parametric in \li`find`, \li`find` itself is implemented
in the module; adding an assumed type declaration would be redundant.
We would rather preserve the existing definition of \li`find`,
and automatically rewrite, e.g., \li`send` into its \li`mk_send` counterpart
that takes \li`find` as an argument. We show in this section how to reach this goal
using metaprogramming.

\paragraph{Rewriting, Formally}
Following the programming pattern described in \sref{no_functor}, we assume
that every function node $g_i$ in the static call-graph is parameterized over an argument
\hl{$\mathsf{idx}:t_\mathsf{idx}$} that represents the specialization index, and that
this argument appears in first position.

At definition site, every function definition
\hl{$\klet f\,\mathsf{idx}\,\overline{x} =$} is
replaced by
\hl{
$\klet \mathsf{mk_f}\,\mathsf{idx}$ $(g_1: t_{g_1}\,\mathsf{idx})$}
\hl{$\ldots (g_n: t_{g_n}\,\mathsf{idx})\,\overline{x} =$}.
The $g_i$ represent all the callees in the body of $f$. 
The $t_{g_i}$ are the types of the original $g_i$,
abstracted over the index
$\mathsf{idx}$, that is, if the type of $g_i$ was the dependent arrow \hl{$\mathsf{idx}:t_\mathsf{idx}
\to t$}, then $t_{g_i}$ is the dependent function
\hl{$t_{g_i} = \lambda (\mathsf{idx}:t_\mathsf{idx}).~t$},
which allows us to write the application $t_{g_i}~\mathsf{idx}$.
At call-site, when encountering a call \hl{$g_i\,\mathsf{idx}\,\overline{e}$},
the call becomes \hl{$g_i\,\overline{e}$} and references the bound variable $g_i$ instead of the global name.

Taking our running example, we have $t_\mathsf{idx} =$ \li+mindex+, and
$\mathsf{eq}: i:t_\mathsf{idx} \to i.k \to i.k \to \mathsf{bool}$. Our goal is
to make sure that \li+find+ becomes parameterized over an argument \li+eq+
specialized for the \emph{same value of the index} as \li+find+. To achieve
that, we pick
$t_\mathsf{eq} = \lambda(i: t_\mathsf{idx}).~i.k \to i.k \to \mathsf{bool}$,
and thus rewrite \li+find+ into
$\klet \mathsf{mk_{find}}~(i: t_\mathsf{idx})~(eq: t_\mathsf{eq}\; \mathsf{idx})$,
which then reduces into 
$\klet \mathsf{mk_{find}}~(i: t_\mathsf{idx})~(eq: i.k \to i.k \to \mathsf{bool})$,
where the index $i$ is the same everywhere, meaning that \li+eq+ is specialized
for the same choice of types as \li+find+.

\paragraph{Recursively Traversing the Call-Graph}

The rewriting presented above is highly modular; it allows us to rewrite each function
in isolation.
Following the same process as for \li`find`, we notice when rewriting \li`send`
that it should be parameterized by a specialized version of \li`find` itself.
Empirically, we observe
the composition of parametric functions to be a common pattern. Instead of manually
applying our rewriting to \li`send`, \li`recv`, and \li`find`, we recursively
traverse the call-graph, automatically performing rewriting on the definitions
of all callees of the function being rewritten. Using this approach, a user
only needs to invoke the rewriting on the API endpoints of their library,
i.e., specific top-level functions.
When encountering a top-level \li`assume` declaration, as described in
\sref{assume:val}, the traversal stops. Callers end up with the
correct additional parameters, and it will be up to the user to exhibit suitable
instantiations for the assumed functions.

\paragraph{Section Variables}
Our mechanism is very similar to the section variables mechanism provided by
provers such as Coq and Lean. For instance, it would be possible to automatically
parameterize \li+find+ with \li+eq+ by using a section in which \li+eq+ is declared as a
\li+Variable+; we show such an example for Coq in Listing~\ref{lst:coqsection}.
The section mechanism in its current shape would however lead to \emph{slightly} more work
on the user side: it would work for all the definitions that we mark as \li+assume+ in
\fstar, as we would simply declare them as section variables, but doesn't provide a
straightforward way of parameterizing functions like \li+send+ and \li+recv+ with
\li+find+.  Indeed, we would need to both define \li+mk_find+ \emph{and} declare \li+find+
as a section variable for \li+send+ and \li+recv+ to use it, so that they get correctly
parameterized; with our call-graph rewriting we write a single definition for \li+find+.

\begin{figure}
\begin{center}
\begin{minted}[linenos,mathescape=true,escapeinside=~~,texcomments]{coq}
Section map.
 Record mindex := { k : Set; v : Set }.
 Variable eq : forall i:mindex, i.(k) -> i.(k) -> bool.
 Definition find (i : mindex) (x : i.(k)) (ls : list (i.(k) * i.(v))) : option i.(v) := ...
End map.
\end{minted}
\end{center}
\captionof{listing}{An equivalent of \li`find` using Coq's section variables}
\label{lst:coqsection}
\end{figure}

\subsection{Fine-Grained Code Specialization}
\label{sec:tactic:eliminate}

While inlining all functions, as explained in \sref{zerocost},
is not desirable, specializing all functions in the call-graph can also
conflict with a programmer's intent. Many functions, e.g.,
\li`alloc` and \li`upd` from the standard library,
are not parametric and thus do not require specialization; furthermore, to reduce the size of
proof contexts and ease verification, programmers often rely on auxiliary functions
that are expected to be inlined at extraction-time.
Consider for instance the while combinator used to implement \li`find`.
While inlining the closures for the loop condition and the loop body
is reasonable for small examples, a programmer might find it useful
to hoist them for verification purposes, as shown in \lref{whilehoist},
while unfolding them at extraction-time to retrieve idiomatic code.

\begin{figure}
\begin{center}
\begin{minted}[linenos,mathescape=true,escapeinside=~~,texcomments]{fstar}
let while_cond (b: pointer bool) (_:unit) = !*b
let while_body (i: mindex) (b: pointer bool) (lsp: list (i.k * i.v)) (x:i.k) (_:unit) =
  let ls = !* lsp in
  match ls with
  | [] -> upd b false | (x', _) :: tl -> if eq x x' then upd b false else upd lsp tl

let find (i : mindex) (x : i.k) (ls : list (i.k * i.v)) : option i.v =
  let b = alloc true in let lsp = alloc ls in
  while (while_cond b) (while_body i b lsp x);
  match !* lsp with | [] -> None | (_, y) :: _ -> Some y
\end{minted}
\end{center}
\captionof{listing}{Hoisting loop closures}
\label{lst:whilehoist}
\end{figure}

Designing generic heuristics to determine which functions should be specialized 
or inlined is tricky; getting them wrong risks alienating developers when they
do not obtain the shape of the code they expect. Instead of a generic
solution, we prefer to leverage programmers'
knowledge of their code. Using \fstar's annotation system, we provide two
attributes, \li`Specialize` and \li`Eliminate`, that enable a fine-grained
control on the rewritings performed by our approach.

Before rewriting, declarations annotated with \li`Eliminate` are preprocessed;
their top-level declarations are removed, and their definitions are inlined
at the different call-sites.\footnote{In practice, we use a more efficient implementation strategy
that allows us to perform our code rewriting in a single pass, and allows us to
avoid traversing big terms that have already undergone a first round of
inlining. The commented implementation of the tactic, in the supplementary
material, contains all of the details.}
After preprocessing, instead of rewriting each function definition and callee
as described in \sref{tactic:rewriting},
we limit the code transformation to functions annotated with the \li`Specialize`
attribute.

We show in \lref{find:attributes} a complete example using the different features
presented in this section.
The code on the right corresponds to the \fstar code on the left, after automatically
rewriting \li`find`. As \li`while_body` and \li`while_cond` are annotated with
\li`Eliminate`, they are inlined during preprocessing.
The \li`eq` function declaration is annotated with the \li`Specialize` attribute;
it therefore appears as an argument to \li`mk_find`. Other functions, i.e.,
\li`alloc` and \li`upd` take their roots in \fstar's standard library,
and are not annotated with any of our custom attributes. They are therefore
ignored and left as-is while statically rewriting the call-graph.
In real developments (see sections \ref{sec:case-studies} and \ref{sec:streaming}),
annotating the functions proved to be extremely straightforward and light.
In return, it allowed us to automatically transform the code into a higher-order version,
which represents a fair amount of work when performed manually.

\af{Code in listings does not fit in margins}
\begin{figure}
\begin{minipage}[t]{0.54\textwidth}
\centering
\begin{minted}[linenos,mathescape=true,escapeinside=~~,texcomments]{fstar}
type mindex = { k : Type; v : Type }

[@ Specialize]
assume val eq (i : mindex) : i.k -> i.k -> bool

[@ Eliminate]
let while_cond (b: pointer bool) (_:unit) = !*b
[@ Eliminate]
let while_body (i: mindex) (b: pointer bool)
  (lsp: list (i.k * i.v)) (x:i.k) (_:unit) =
  ... (* elided, same as \lref{whilehoist} *)

[@ Specialize] let find (i : mindex) (x : i.k)
  (ls : list (i.k * i.v)) : option i.v =
  let b = alloc true in let lsp = alloc ls in
  while (while_cond b) (while_body i b lsp x);
  match !* lsp with
  | [] -> None | (_, y) :: _ -> Some y
\end{minted}
\end{minipage}
\begin{minipage}[t]{0.44\textwidth}
\centering
\begin{minted}[linenos,mathescape=true,escapeinside=~~]{fstar}
(* After rewriting *)
type mindex = { k : Type; v : Type }

let mk_find (i: mindex) (eq: i.k-> i.k -> bool)
  (x: i.k) (ls: list (i.k * i.v)) : option i.v =
  let b = alloc true in let lsp = alloc ls in
  while (fun () -> !* b)
    (fun () -> let ls = !* lsp in
     match ls with
     | [] -> upd b false
     | (x', _) :: tl ->
      if eq x x' then upd b false
      else upd lsp tl);
  match !* lsp with 
  | [] -> None
  | (_, y) :: _ -> Some y
\end{minted}
\end{minipage}
\captionof{listing}{Fine-grained control on code transformation}
\label{lst:find:attributes}
\end{figure}

\subsection{Implementation in Meta-\fstar}

We have implemented this call-graph rewriting using syntax inspection, term
generation and definition splicing in Meta-\fstar~\cite{fstar-meta}.
Meta-\fstar allows the programmer to script the \fstar compiler using
\emph{user-written \fstar programs}, a technique known as elaborator reflection
and pioneered by Lean~\cite{lean2015} and Idris~\cite{idris}.
This approach means that any fresh term generated by a meta-program
must be re-checked for soundness; we therefore do not prove any results
about our procedure and let \fstar validate the terms we produce.
When calling our procedure, the user passes the roots of the call-graph
traversal, i.e., the API endpoints of their library,
along with the type of the index. The procedure traverses the
call-graph, generates rewritten variants of all the definitions, and inserts
them at the current program point.

We insist on the fact that the entire rewriting procedure was implemented in user-space
and does not need to be trusted. Of course, one might wonder how we enforce that
the types of the generated, higher-order definitions are correct.
Indeed, if our meta-program generates definitions which don't have the correct type,
successively type-checking those definitions against those types doesn't give us any guarantees.
In practice, we check this later when \emph{instantiating} those higher-order definitions,
by annotating their specializations: if the types generated by our meta-program were
incorrect, type checking would fail on this part of the code. As we use helpers to factor out
types between the generic and the specialized definitions, annotating those instantiations
doesn't create any burden on the user side.
Finally, one last point of concern would be that our rewriting procedure transforms the
functions in such a way that the generated C code has an unexpectedly poor performance.
We note that, due to the nature of the transformations we perform, which consist, after
instantiation and partial evaluation, in specializing part of code, this shouldn't happen
in practice. Of course, this doesn't dispense us from benchmarking the code, which we do.
As our technique gives users fine-grained control on the shape of the generated code,
it is also possible to tune the output to reach the desired performance.
In particular, we did not see any noticeable change of performance after adapting the code
from \haclstar to use the present technique (\sref{case-studies}).
The interested reader can find our entire, generously commented implementation
in the file \li`Meta.Interface.fst` of the supplementary material.

\paragraph{Comparison with existing techniques}
Our mechanism shares similarities with other type specialization techniques.
Specifically, Haskell’s \li+SPECIALIZE+ pragma and Rust’s trait system attempt to solve
attempt to solve very similar problems, albeit as a trusted whole-program monomorphization
pass within their respective compilers, as opposed to a source-to-source rewriting pass.
Putting aside the problem of working within a proof assisant, we note that by contrast our
technique is 1) untrusted and thus doesn’t require extending the F* compiler, 2) allows
specializing over values, functions, while leveraging general-purpose dependent types, and
3) gives the user fine-grained control on how the specialized call-graph should look
like, in particular for the purpose of outputting a readable program.

\paragraph{Adaptability of our technique to other proof assistants.}
While we implemented our approach in \fstar, our techniques are not tied to one particular language.
Our work focuses on the verification of shallowly embedded programs;
although a discussion of the advantages and disadvantages of the use of deep embeddings or
shallow embeddings is out of scope of this paper, it is worth noting that the
extraction of shallowly embedded programs has been used in many other verification
projects, relying on a variety of proof
assistants~\cite{compcert,cakeML,fiat-crypto,FiatToFacade,rupicola,lammich2019refinement}.
Restricting our scope to the verification of shallowly embedded programs, our approach
needs the following key ingredients to be applicable. (1) We need to be able to encode
functors, and as we explained in \sref{background} there exists a well known
technique to do so in dependently-typed languages such as Coq, Lean or Idris. (2)
We need elaborator reflection to implement the rewriting procedure; some languages like
Lean or Idris provide it in their meta-language, while some other tools like Coq would
require writing a plugin. (3) We need the ability to partially evaluate the specialized
programs, which is a common feature of the aforementionned tools. (4) We need
an extraction mechanism, which is supported for instance by Coq, Lean and Idris;
in particular, we note that Lean and Idris support the extraction to a low-level language
such as C or C++. We thus conclude that our methodology could be ported to either one of these
proof assistants without fundamental difficulties.

\section{Application to the \haclstar Cryptographic Library}
\label{sec:case-studies}

We introduced our approach on a small example in the previous sections.
We now demonstrate its applicability on real-world examples by presenting its
use on heavily optimized implementations of cryptographic
primitives inherited from the \haclstar~\cite{hacl,haclxn} and
EverCrypt~\cite{evercrypt} libraries.
\haclstar is a cryptographic library written in \fstar which compiles to C;
it offers vectorized versions of many algorithms via C compiler intrinsics,
e.g., for targets that support AVX, AVX2 or ARM Neon.
EverCrypt is a high-level API that multiplexes between \haclstar and
Vale-Crypto~\cite{vale,vale-fstar}, a library of verified primitives implemented in
assembly; it supports dynamic selection of algorithms and
implementations based on the target CPU's feature set.
Combined with EverCrypt, \haclstar features 105k lines of F* code for 72k lines
of generated C code (excluding comments and whitespace, as well as the Vale assembly DSL).
Those case studies are not new, but were adapted to apply our technique.
We explain in this section how we achieved this, and by doing so show how they stress
all the requirements which motivated our new
approach and which we described in the past sections;
that is, the need for 1. zero-cost abstractions which provide high-level
modularity and composability; 2. a fine-grained control on the shape of the
generated code to obtain efficient and idiomatic implementations;
3. a flexible and lightweight approach which limits the amount of boilerplate
and handles a wide range of scenarios.
We detail in \sref{evaluation:core-algos} the limitations of the previous techniques,
and the consequent benefits we got by applying our new approach.

\subsection{Generically Writing Hardware-Specialized Code: ChaCha20-Poly1305}
\label{sec:chacha}

We first present the application of our approach on one of our simplest
examples, the ChaCha20-Poly1305 cryptographic construction.
This case study illustrates how we used our approach to generate, from a single
generic implementation, optimized code specialized for specific hardware
targets.
ChaCha20-Poly1305 is an algorithm for authenticated encryption with additional data
(AEAD). The specifics of the construction are orthogonal to this paper; for
presentation purposes, it is sufficient to know that it combines two
cryptographic primitives: the ChaCha20 stream cipher, and the Poly1305 message
authentication code (MAC).

Depending on the hardware used, both ChaCha20 and Poly1305
admit several implementations. In particular, these primitives are especially well-suited to
SIMD vectorization, by which we apply an operation (e.g., multiply by a
constant) on all the elements of a vector at the same time, and can be highly
optimized when such instructions are available.
Previous work on \haclstar~\cite{haclxn} demonstrated how to write and verify generic,
vectorization-agnostic implementations
of these algorithms, which could be specialized by partial evaluation to provide idiomatic
C implementations.
The approach then used to make the implementation generic was plagued with various
issues, whose detailed description we defer until the \sref{evaluation};
for now, it suffices to say that it struggled with scalability.

We now show how we implemented the cryptographic construction in our DSL. 
We mentioned earlier (\sref{no_functor}) that the index captures the set of
possible specializations. Our running example admitted an infinite set of possible
specialization choices, as long as the key type admitted a decidable equality. In the
example below, we only capture a \emph{finite} set of possible specialization choices,
which we express via a finite enumeration of type \li`arch_index`.

\begin{minted}{fstar}
type arch_index = | V32 | V128 | V256

[@ Specialize ] assume val chacha20_encrypt: w:arch_index -> chacha20_encrypt_st w
[@ Specialize ] assume val do_poly1305: w:arch_index -> poly1305_st w

let aead_encrypt (w:arch_index) ... =
  chacha20_encrypt w ...;
  do_poly1305 w ... 
\end{minted}

To parameterize over both primitives, we rely on abstract signatures for
ChaCha20 and Poly1305, as described in \sref{assume:val}.
The types \li`chacha20_encrypt_st` and \li`poly1305_st` correspond to the
function types of both primitives, where the type of the arguments (e.g., the
Poly1305 context) depend on the \li`w: arch_index` parameter. Both functions are
annotated with the \li`Specialize` attribute, indicating that they are
parameters of the implementation.
As \li`aead_encrypt`
calls these two functions, our rewriting procedures generates a higher-order
combinator \li`mk_aead_encrypt` which requires two functions for
\li`chacha20_encrypt` and \li`do_poly1305`.
The \li`aead_decrypt` function is rewritten in a similar manner.
The last step is to instantiate this combinator with different existing implementations for both
primitives, for instance one specialized for 128-bit vectorization.

\begin{minted}{fstar}
let aead_encrypt : aead_encrypt_st V128 = mk_aead_encrypt V128 do_poly1305_128 chacha20_encrypt_128
\end{minted}

The resulting C code is idiomatic, and close to what one would expect from handwritten C code, albeit with
formal guarantees about its correctness and constant-time execution.
Case in point, the corresponding
code in the \haclstar library was previously integrated into Mozilla Firefox~\cite{haclxn}.

\subsection{Composing Implementations: Curve25519}
\label{sec:curve}

We saw in the previous section a first application of the basic features of our
approach. In this section, we demonstrate how our technique gives us
\emph{composability} on a real-world example, allowing us to simplify a
collection of verified implementations of a widely used elliptic curve,
Curve25519~\cite{curve25519}.

The specifics of the algorithm are out of scope for this paper; in this
presentation, it suffices to say that it relies on modular arithmetic in a
mathematical field, which admits two implementations based on different
representations of the field elements. Furthermore, one of these representations
relies on a set of primitives (e.g., addition)
which themselves admit two different implementations, one in \lowstar, and one
in Vale assembly when specific hardware instructions are available.

Previous work on EverCrypt~\cite{evercrypt} provided a single verified
client-facing API multiplexing between different implementation, that is, an API
which selects the best implementation depending on the hardware available;
these implementations however lived side by side, duplicating a lot of code.
Using our approach, we now show how we reduce code redundancy, by aggressively
sharing more code between those different implementations, and only specializing
between the different field representations and implementations \emph{a
posteriori}.
Providing a single generic implementation that will be automatically specialized
reduces the maintenance cost of the \haclstar codebase, while also simplifying
the development of algorithmic improvements across our different versions.
Using OCaml syntax, the end result allows users to pick between three different versions of Curve25519:
\li+module Curve64Lowstar = Curve25519(Field64(CoreLowstar))+,
\li+module Curve64Vale = Curve25519(Field64(CoreVale))+, and
\li+module Curve51 = Curve25519(Field51)+.
An important point to notice is that we leverage our DSL to organize our
implementation into three layers, that we later compose with each other.
For presentation purposes, we present here a simplified version of Curve25519
which omits several layers and functions.  We refer the interested reader to the
supplementary material for our complete implementation.

\paragraph{Composing Abstractions.}

Curve25519 exposes several functionalities, including the function \li`scalarmult`, which
performs scalar multiplication on the elliptic curve.
This function calls into \li`encode_point`, which itself relies on the field addition
\li`fadd`. All these functions are parameterized by an index corresponding to the
field representation, of type \li`field_index`. For clarity of the generated code,
we wish to avoid inlining any of these functions; we thus annotate each
definition with the \li`Specialize` attribute.

\begin{minted}{fstar}
type field_index = | F51 | F64

[@ Specialize ] assume val fadd (i:field_index) -> fadd_t i

[@ Specialize ] let encode_point (i: field_index) ... = ... fadd ...
[@ Specialize ] let scalarmult (i:field_index) ... = ... encode_point ...
\end{minted}

The code is rewritten as one might expect. We can provide \emph{multiple} specializations
for one choice of index. If \li+i+ is \li+F64+, we can generate both
\li+encode_point_64_lowstar+ and \li+encode_point_64_vale+:

\begin{minted}{fstar}
let encode_point_64_lowstar = mk_encode_point F64 Lowstar.Field64.fadd ...
let encode_point_64_vale = mk_encode_point F64 Vale.Field64.fadd ...
\end{minted}

In addition, we can generate \li+encode_point_51+ (elided). This in turns allows us to generate
three versions for \li+scalarmult+:

\begin{minted}{fstar}
let scalarmult_51 = mk_scalamult F51 encode_point_51 ...
let scalarmult_64_lowstar = mk_scalamult F64 encode_point_64_lowstar ...
let scalarmult_64_vale = mk_scalamult F64 encode_point_64_vale ...
\end{minted}

In effect, leveraging the composability permitted by our approach,
the Curve25519 implementation is organized into three layers:
the field arithmetic, the field encoding (\li+F51+ or \li+F64+), and the elliptic curve
operations.

\subsection{A Highly Parametric Example: The HPKE Construction}

We now present the culmination point of our series of cryptographic
primitives: HPKE~\cite{hpke} (Hybrid Public-Key Encryption), a recent
cryptographic construction that combines AEAD (Authenticated Encryption with
Additional Data), DH (Diffie-Hellman), and hashing.
The implementation of HPKE ticks several of the boxes that we wished to cover
with our technique, that is: we build on top of several functionalities, each of these 
functionalities can be instantiated with several algorithms (e.g., Curve25519 or P256 for DH,
ChaCha20-Poly1305 or AES-GCM for AEAD), and every algorithm admits several implementations;
we have a complex call graph divided into several layers.
Omitting several definitions for brevity, we structure the code as follows, using \li`hpke_alg` as our index.

\begin{minted}{fstar}
type aead_alg = AES128_GCM | AES256_GCM | CHACHA20_POLY1305
type hpke_alg = dh_alg * aead_alg * hash_alg

type key_aead (alg: hpke_alg) = lbuffer U8.t (key_len (snd3 alg))

[@ Specialize ] assume val sign: (alg:hpke_alg) -> sign_t alg
[@ Specialize ] assume val enc: (alg:hpke_alg) -> enc_t alg

[@ Eliminate ] let helper (alg: hpke_alg): helper_t alg = fun ...  -> ... sign alg ...
[@ Specialize ] let hpke_sealBase (alg: hpke_alg): hpke_sealBase_t alg = fun ... ->
    ... helper alg ...
    ... enc alg ...
\end{minted}

The index \li`hpke_alg` is a triple that captures all possible algorithm
choices prescribed by the HPKE RFC. We thus write specifications, lemmas,
helpers, and types parametrically over the index as standalone definitions. The
\li`key_aead` type, for example, is parametric over triplets of algorithms, and
defines a low-level key to be an array of bytes whose length is the key length
for the chosen AEAD. The same systematic parameterization over \li`hpke_alg` can be
carried to functions and their types, e.g., \li`hpke_sealBase`, which encrypts and
authenticates a plaintext.
We use small helpers, e.g., \li+helper+, to make verification robust in the
presence of an SMT solver and ensure modularity of the proofs, as explained in
\sref{tactic:eliminate}; because we want to evaluate them away at extraction
time, we mark them with the \li+Eliminate+ attribute.

One possible specialization, out of hundreds of possible options, is
to pick the \li`F51` version Curve25519 for DH (\sref{curve}), the AVX 128-bit variant of
Chacha20-Poly1305 for AEAD (\sref{chacha}), and SHA2-256 for hashing.

\begin{minted}{fstar}
let alg = (DH_CURVE25519, CHACHA20_POLY1305, SHA2_256)
let sealBase = mk_hpke_sealBase alg aead_encrypt_cp128 scalarmult_51 ...
\end{minted}

The HPKE example is emblematic of our modularity pattern.  It allows the programmer to
author their verified code while thinking about the choice of \emph{functionalities};
picking concrete implementations for each functionality and specializing the code
accordingly is left to a later phase, and is entirely handled by our automated rewriting. All the user
has to do is pick their particular choice of algorithms and implementations, and enjoy the resulting
specialized HPKE.

Out of hundreds of possible choices, the \haclstar library provides 30
different variants of HPKE.
Adding a new variant requires minimal effort; furthermore, with our methodology,
each variant lives in its own
separate file, which can then be compiled with exactly the right compiler
options without any danger of miscompilation.

\section{A Generic State Machine: the Streaming API}
\label{sec:streaming}

In the previous section, our technical contributions consisted of honing the
proofs and restructuring the codebase of \emph{pre-existing} algorithms, solving
deep technical roadblocks in the process.
In this section, we
describe a novel case study that was enabled by the present work. We first explain the nature of the
problem; then show how our methodology came in judiciously and allowed us to structure
our code to achieve maximum modularity.

We want to emphasize that this case study is an important contribution,
\emph{on its own}, for two reasons. First,
it encompasses all the difficulties of carrying out large-scale verification of
low-level code: the development is built on top of
already complex implementations (i.e., the \haclstar hashes); it is divided into several
modular layers, which must each be specialized in a myriad of ways; finally,
unverified implementations of this code have historically caused critical bugs in
high-profile software~\cite{sha3-bug,mouha2018finding},
and this complexity pervades our invariants, which were subtle and
difficult to get right. Second, and perhaps more importantly: the
cryptographic community has some folk knowledge of what a block algorithm is;
but as far as we know, this folk knowledge was never distilled into formal,
precise language, like we do here.

\subsection{Illustrating Streaming APIs with the Hash example}

Many cryptographic algorithms offer identical or similar \emph{functionalities}. For example, SHA2~\cite{sha2}, SHA3~\cite{sha3}, and Blake2~\cite{blake2-paper,blake2}
(in no-key mode) all implement the \emph{hash} functionality, taking an input text to compute a resulting
digest.
As another example, HMAC~\cite{hmac}, Poly1305~\cite{poly1305}, GCM~\cite{gcm}, and Blake2 implement the \emph{message authentication code (MAC)}
functionality, taking an input text and a key to compute a digest.

At a high level, these functionalities are simply black boxes with one or two inputs, and a single output.
Taking \haclstar's SHA2-256 implementation as an example, this results in a natural, self-explanatory C API:

\begin{minted}{c}
void sha2_256(uint8_t *input, uint32_t input_len, uint8_t *dst);
\end{minted}

This ``one-shot'' API, however, places unrealistic expectations on clients of this
library. For instance, the TLS protocol, widely-used to secure internet communications,
computes repeated intermediary hashes of the handshake data transmitted so far. Using
the one-shot API would be grossly inefficient, as it would require re-hashing
the entire handshake data every single time. In other situations, merely hashing
the concatenation of two non-contiguous arrays with this API requires a full
copy into a contiguous array.

Cryptographic libraries thus need to provide a different API that allows clients to perform
\emph{incremental} hash computations. A natural candidate for this is the block API:
all of the algorithms we mentioned above are block-based, meaning that, under the hood, they
follow the state machine from \fref{block-machine}: after allocating an internal state
(\li`alloc`), they initialize it (\li`init`), process the data (\li`update_block`)
block by block (for an algorithm-specific block size), perform some special treatment
for the leftover data (\li`update_last`), then extract the internal state (\li`finish`) onto
a user-provided destination buffer, which then holds the final digest.
Revealing this API allows clients to feed their data into the hash
\emph{incrementally}, meaning that at first glance, our earlier issues are
solved as we have found a way to hash data block by block without holding onto
the entire input.

\begin{figure}
  \begin{minipage}[t]{0.57\textwidth}
  \centering
  \begin{tikzpicture}[state/.style={
    draw,thick,circle,minimum height=2em,node distance=4.5em
    },label/.style={
      font=\sffamily\footnotesize
    }]
  \node [] (create) { };
  \node [state, right of = create] (init) { };
  \node [state, right of = init] (state1) { };
  \node [state, right of = state1] (state2) { };
  \node [state, right of = state2] (state3) { };
  \node [right of = state3,node distance=4em] (done) { };

  \draw [ultra thick,->] (create) --
    node [above,label] { alloc }
    (init);

  \draw [ultra thick,->] (init) --
    node [above,label] { init }
    (state1);

  \draw [ultra thick,->] (state1) --
  node [above,label,text width=2em,align=center] { update last }
    (state2);

  \draw [ultra thick,->] (state2) --
    node [above,label] { finish }
    (state3);

  \draw [ultra thick,->] (state3) --
    node [above,label] { free }
    (done);

  \path [ultra thick,->]
    (state1)
    edge [loop below]
    node [label] { update block }
    (state1);

  \draw [ultra thick,->] (state2)
    to [out=90,in=90]
    node [above,label] { \quad init }
    (state1);

  \draw [ultra thick,->] (state3)
    to [out=90,in=90]
    node [above,label] { init }
    (state1);

  \end{tikzpicture}
  \centering
  \caption{State machine of an error-prone block-based API}
  \label{fig:block-machine}
  \end{minipage}
  \begin{minipage}[t]{0.42\textwidth}
  \begin{tikzpicture}[state/.style={
    draw,thick,circle,minimum height=2em,node distance=4em
    },label/.style={
      font=\sffamily\footnotesize
    }]
  \node [] (init) { };
  \node [state, right of = init] (state1) { };
  \node [right of = state1,node distance=4em] (done) { };

  \draw [ultra thick,->] (init) --
    node [above,label] { alloc }
    (state1);

  \draw [ultra thick,->] (state1) --
    node [above,label] { free }
    (done);

  \path [ultra thick,->, draw]
    (state1)
    to [ out=-110, in=-145, looseness = 10 ]
    node [label, below] { reinit }
    (state1);

  \path [ultra thick,->, draw]
    (state1)
    to [ out=-35, in=-70, looseness = 10 ]
    node [label, below] { update }
    (state1);

  \path [ultra thick,->, draw]
    (state1)
    to [out = 70, in = 110, looseness = 8]
    node [label,above] { finish }
    (state1);

  \end{tikzpicture}
  \centering
  \caption{State machine of a safe, streaming API}
  \label{fig:safe-machine}
  \end{minipage}
\end{figure}

The issue with this block API is that it is wildly unsafe to call from unverified C code.
First, it requires clients to maintain a block-sized buffer that, once full, must be
emptied via a call to \li`update_block`. This entails non-trivial modulo-arithmetic computations
and pointer manipulations, which are error-prone~\cite{mouha2018finding,sha3-bug}.
Second, clients can easily violate the state machine. For instance, when extracting
an intermediary hash, clients must remember to \li`copy` the internal hash state,
call the sequence \li`update_last` and \li`finish` on the copy, free that copy, and only
then resume feeding more data into the original hash state.
Third, algorithms exhibit subtle differences: for instance, Blake2 must not receive empty
data for \li`update_last`, while SHA2 does not suffer from this restriction.
In short, the block API is error-prone, confusing, and is likely to result in programmer
mistakes.

We thus wish to take all of the block-based algorithms, and devise a way to wrap their
respective block APIs into a uniform, safe API that eliminates all of the pitfalls above.
We dub this safe API the streaming API (\fref{safe-machine}): it has a degenerate state
machine with a single state; it performs buffer management under the hood; it hides the
differences between algorithms; and performs necessary copies as-needed when a digest
needs to be extracted.

Writing and verifying a copy of the streaming API for each one of the eligible algorithms
would be tedious, not very much fun, and bad proof engineering. Instead, we apply the
methodology exposed throughout this paper, and set out to write a generic
API transformer that turns any block algorithm into its safe, streaming
counterpart. We begin with a description of a block algorithm's stateful API and
intended specification using our DSL -- this will be our ``functor argument''.

\subsection{The Essence of Stateful Data}
Before we get to the block API itself, we need to capture a more basic notion,
that of an abstract piece of data that lives in memory, composes with the \lowstar
memory model and modifies-clause theory~\cite{kassios2006dynamic},
and supports basic operations such as allocation, de-allocation, and copy.
This is the \emph{stateful} API presented in \lref{stateful}.

\begin{figure}[t]
  \centering
  \begin{minted}[linenos]{fstar}
type state_index = {
  s : Type; (* Low-level type *)
  t : Type; (* A pure representation of an s *)
  footprint : mem -> s -> Ghost loc;
  invariant : mem -> s -> Ghost Type;
  v : mem -> s -> Ghost t; (* Reflect an s in a memory snapshot as a pure value *)

  (* Adequate framing lemmas *)
  frame_invariant: ss:state_index -> l:loc -> s:ss.s -> h0:mem -> h1:mem -> Lemma
    (requires (ss.invariant h0 s /\ loc_disjoint l (ss.footprint h0 s) /\ modifies l h0 h1))
    (ensures (ss.invariant h1 s /\ ss.v h0 s == ss.v h1 s /\ ss.footprint h1 s == ss.footprint h0 s))
  ... (* Omitted: additional lemmas *) }

(* Stateful operations *)
[@Specialize] assume val malloc (#i: state_index) ... : ST ...
[@Specialize] assume val free (#i: state_index) ... : ST ...
[@Specialize] assume val copy (#i: state_index) ... : ST ...
... (* Omitted *)
  \end{minted}
  \captionof{listing}{The \li+stateful+ API}
  \label{lst:stateful}
\end{figure}

We parameterize the implementation with the record \li`state_index`. We
mentioned earlier that the index captures the space of all possible instantiations
-- here, this space is constrained by the presence of valid specifications that
satisfy the behavioral lemmas we require. This is an extension of our previous
style: the index bundles up in one record all of the type-level arguments to
our functions.
The specifications are used only in the proofs and are not relevant at runtime; for this
reason we put them in the index and mark them as \li+ghost+ by using the \li+Ghost+ effect. 
Similarly to frameworks like Why3 or Dafny, ghost code (and variables) in
\fstar is computationally irrelevant code; as such it must obey some restrictions, for
instance, it is possible to convert a non-ghost value to a ghost value, but not the other
way around. At extraction time, ghost code is erased, typically by being replaced with
unit values (which are later eliminated).
As the specifications are grouped in the index, they also do not undergo code specialization
and higher-order rewriting, and do not need to be annotated with our DSL.

This establishes a distinction between erased arguments (types, specifications,
lemmas), which are handled via regular polymorphism and as such appear within
the index, and run-time functions, which must undergo rewriting, higher-order
parameterization, and as such rely on \li+assume val+ and our rewriting
mechanism.

Importantly, we saw in \sref{case-studies} the use of closed enumeration types
for the choice of the index, by which we allow a \emph{finite} set of possible
specializations. In the present case, due to the highly generic nature of our code we
need to use an open ended parameterization (i.e., a record), by which the index captures
an \emph{infinite} set of possible choices of specialization.

The \li`state_index` record contains a low-level type \li`s` (e.g., \li`lbuffer U8.t 64ul`, an array of length 64 containing bytes)
which comes with an abstract \li`footprint` (e.g., the extent of that array in memory), and an
abstract \li`invariant` (e.g., the array is \li`live`).
The \li`footprint` and the \li`invariant` live in the \li`Ghost` effect, meaning they are
computationally irrelevant and thus erased at extraction.
The low-level type
can be reflected as a pure value of type \li`t` (e.g., a sequence) using a
ghost function \li`v` (e.g., \li`as_seq`, which interprets arrays as pure sequences).
Outside of \li`state_index`, we declare some administrative lemmas which
allow harmonious interaction with \lowstar's modifies-clause
theory;
for instance, the \li`frame_invariant` lemma which we need because of the specificities of
the \lowstar memory model:
under the pre-condition
that the state invariant holds in an initial memory snapshot \li`h0`, and that the memory
locations modified between \li`h0` and \li`h1` are disjoint from the state footprint, then
the invariant also holds in \li`h1` and the (pure reflection of the) state and the
footprint are left unchanged; we automate its application with an SMT pattern
(elided), which
indicates to Z3 when to instantiate this lemma.
The stateful operations allow, respectively, allocating a fresh state on
the heap; freeing a heap-allocated state; and copying the state.

As we need two different stateful objects for our block implementation,
states and keys (see~\ref{sec:block}), we actually declare two stateful APIs,
in modules \li`State` and \li`Key` respectively;
note that in practice we factor out the types of the
declarations, so as not to duplicate code.
Writing instances of the stateful APIs is easy, the most complex one being the
internal state of Blake2 which occupies 46 lines of code,
with all proofs going through automatically.

\subsection{The Essence of Block Algorithms}
\label{sec:block}

\begin{figure}[t]
  \centering
  \begin{minted}[linenos,escapeinside=~~]{fstar}
type block_index = {
  km: key_management; (* km = Runtime \/ km = Erased *)
  state: State.state_index; (* State spec *)
  key: Key.state_index; (* Key spec *)
  
  (* Some lengths used in the specs *)
  max_input : x:nat { 0 < x /\ x < pow2 64 };
  out_len: x:U32.t { U32.v x > 0 };
  block_len: x:U32.t { U32.v x > 0 };

  (* The one-shot specification *)
  spec_s : key.t -> input:seq U8.t{length input <= max_input}) -> out:seq U8.t{length out == U32.v out_len};

  (* The block specification *)
  init_s : key.t -> state.t; ~\label{line:block:spec_beg}~
  update_multi_s : state.t -> prevlen:nat -> s:seq U8.t{length s %
  update_last_s : state.t -> prevlen:nat -> s:seq U8.t{length s <= U32.v block_len} -> state.t;
  finish_s : key.t -> state.t -> s:seq U8.t { length s = U32.v out_len };~\label{line:block:spec_end}~ 

  update_multi_is_a_fold : ... -> Lemma ...; (* update_multi_s respects the fold law *)~\label{line:block:fold}~
  
  (* Central correctness lemma of a block algorithm *)
  spec_is_incremental: key:key.t -> input:seq U8.t{length input <= max_input} -> Lemma (
    let bs, l = split_at_last (U32.v block_len) input in
    let hash0 = update_multi_s (init_s key) 0 bs in
    let hash1 = finish_s key (update_last_s hash0 (length bs) l) in
    hash1 == spec_s key input); }

[@Specialize] assume val update_multi (bi:block_index) (s:bi.state.s) (prevlen:U64.t) ~\label{lst:block:update_multi}~
  (blocks:buffer U8.t { length blocks %
  (len: U32.t { U32.v len = length blocks /\ ... (* omitted *) })
  ST unit (requires ... (* omitted *))
    (ensures (fun h0 _ h1 ->
      modifies (bi.state.footprint h0 s) h0 h1 /\ ~\label{lst:block:upd:modifies}~
      bi.state.footprint h0 s == bi.state.footprint h1 s /\ ~\label{lst:block:upd:footprint}~
      bi.state.invariant h1 s /\ ~\label{lst:block:upd:inv}~
      bi.state.v i h1 s == bi.update_multi_s (bi.state.v i h0 s) (U64.v prevlen) (as_seq h0 blocks) /\ ...)) ~\label{lst:block:upd:spec}~

... (* Omitted: rest of the block API, e.g. init, finish... *)
  \end{minted}
  \captionof{listing}{The \li+block+ API}
  \label{lst:block}
\end{figure}

We now capture the essence of a block algorithm by authoring an API that
encapsulates a block algorithm's types, representations, specifications, lemmas,
and stateful implementations in one go.
We need the \li`block` API to capture four broad traits of a block algorithm,
namely i) explain the runtime representation and spatial characteristics of the block
algorithm, ii) specify as pure functions the transitions in the state machine, iii) reveal
the block algorithm's central lemma, i.e., processing the input data block by block is the same as processing all of the data in one go, and iv) expose the low-level run-time functions that
realize the transitions in the state machine. The result appears in \lref{block};
for conciseness, we omit the full statement of the fold lemma, as well as the stateful
type of the remaining transitions of the state machine.
Similarly to the stateful API, we gather the specification of the block API in the
index, that is in the record \li`state_index`.
The actual definition is about 150 lines of \fstar, and appears in the anonymous
supplement.

\paragraph{Run-time characteristics.}
A block algorithm revolves around its state, which implements the
\li`State` stateful API.
It may need to keep a key at run-time
(\li`km = Runtime`, e.g., Poly1305), 
or keep a ghost key for specification purposes (\li`km = Erased`, e.g., keyed Blake2),
or may need no key at all, in which case the \li`key` field is a degenerate instance of
the \li`Key` stateful API, such that \li`key.s = unit`.

\paragraph{Specification.}
Using \li`state.t`, i.e., the algorithm's state reflected as pure value, we specify each
transition of the state machine at
lines~\ref{line:block:spec_beg}-\ref{line:block:spec_end}.
Importantly, rather than specify
an ``update block'' function, we use an ``update multi'' function that can process
multiple blocks at a time.
We do not impose any constraints on how \li`update_multi` is
authored, we only request that it obeys the fold law 
\li`update_multi_s ((update_multi_s s l1 b1) (l1 + length b1) b2) == update_multi_s s l1 (concat b1 b2)`
via the lemma
\li`update_multi_is_a_fold` (line~\ref{line:block:fold}).

This style has several advantages. First, this leaves the possibility for optimized algorithms
that process multiple blocks at a time to provide their own \li`update_multi` function, rather
than being forced to inefficiently process a single block. For unoptimized algorithms that
are authored with a stateful \li`update_block`, we provide a higher-order combinator that
derives an \li`update_multi` function and its correctness lemma automatically.
Second, by abstracting over how blocks are processed, we capture a wide range of
behaviors.
For instance, Poly1305 has immutable internal state for storing
precomputations, along with an accumulator that changes with each call to \li`update_block`:
we simply pick \li`state.t` to be a pair, where the fold only operates on the second component.

\paragraph{The block lemma.}
The \li`spec_is_incremental` lemma captures the key correctness condition and ties all of
the specification functions together; by doing so it also specifies the order of the
transitions of the state machine. For a given piece of data, the result \li`hash1`,
obtained via the incremental state machine from \fref{block-machine}, is the same as
calling the one-shot specification \li`spec_s`. This lemma relies on a helper,
\li`split_at_last`, which splits a sequence into a series of blocks and a rest,
and was carefully crafted to subsume the different behaviors between Blake2 and
other block algorithms; in particular, it makes sure the rest is not empty
unless the initial sequence is empty, so that \li`update_last` is never called
on an empty sequence in the case of Blake2.

\begin{minted}{fstar}
let split_at_last (block_len: U32.t) (b:seq U8.t) =
  let n = length b / block_len in
  let rem = length b %
  let n = if rem = 0 && n > 0 then n-1 else n in
  let blocks, rest = split b (n * l) in blocks, rest
\end{minted}

\paragraph{Stateful implementations.}
We now zoom in on the \li`update_multi` low-level signature, which describes a block's algorithm
run-time processing of multiple blocks in one go
(\lref{block}).
This function is characterized by the spec-level \li`update_multi_s`;
under the proper preconditions (elided here),
it only affects the memory locations of the state \li`s` (line~\ref{lst:block:upd:modifies}),
preserves the footprint (line~\ref{lst:block:upd:footprint}) and the invariant
(line~\ref{lst:block:upd:inv}),
and updates the state according to the pure spec (line~\ref{lst:block:upd:spec}).

The combination of \li`spec_is_incremental` along with the \lowstar signatures of
\li`update_multi` and others restricts the API in a way that the only valid usage is dictated
by \fref{block-machine}.
Designing this API while looking at a wide range of
algorithms forced us to come up with a precise, yet general enough description of what a block
algorithm is. We have been able to author instances of this API for SHA3, SHA2 (4 variants),
Blake2 (4 variants), Poly1305 (3 variants), and legacy algorithms MD5 and SHA1.
This includes the vectorized variants of these algorithms, when available.
By materializing those instances, we were able to tie together a whole class of algorithms
under a single unifying interface, therefore materializing the (informal) claim from the
cryptographic community that ``these are all blocks algorithms''.

\subsection{A Streaming API}
Equipped with an accurate and precise description of what a block algorithm is,
we are now ready to use our approach to write an API transformer that takes an instance
of the block API, implementing the state machine from \fref{block-machine}, and returns
the safe API from \fref{safe-machine}.
We now present the definition of the run-time state of streaming API.
The state is naturally parameterized over a \li`block_index`, and wraps the block algorithm's
state with several other fields.

\begin{minted}{fstar}
[@ CAbstractStruct]
type state_s (bi: block_index) = {
  block_state: bi.state.s;
  buf: buffer U8.t { length buf = bi.block_len };
  total_len: U64.t;
  seen: erased (seq U8.t);
  p_key: optional_key bi.km bi.key; }

let state (bi: block_index) = pointer (state_s bi)
\end{minted}

The \li`CAbstractStruct` attribute ensures that the C code below will appear in the
header. This pattern is known as "C abstract structs" and is commonly used by C
programmers to provide a modicum of abstraction: the client cannot allocate
structs or inspect private state, since the definition of the type is not known; it can
only hold pointers to that state, which forces them to go through the API.

\begin{minted}{c}
struct state_s;
typedef struct state_s *state;
\end{minted}

First, \li`buf` is a block-sized internal buffer, which relieves the client of having to
perform modulo computations and buffer management.
Once the buffer is full, the streaming API calls the underlying block algorithm's
\li`update_multi` function, which effectively folds the blocks into the \li`block_state`.
The \li`total_len` field keeps track of how much data has been fed so far, information that
is needed for many block-based algorithms, notably hashes which encode the length of the
input as part of the final block in order to rule out padding attacks.

The most subtle point is the use of a ghost (i.e., computationally irrelevant and
erased at extraction time) sequence of bytes, \li`seen`, which keeps track of the past, i.e.,
the bytes we have fed so far into the hash. This is reflected in the invariant,
which states that if we split the input data into blocks, then the current block algorithm
state is the result of accumulating all the blocks into the block state;
the rest of the data that doesn't form a full block is stored in \li`buf`.

\begin{minted}{fstar}
let state_invariant (bi: block_index) (h:mem) (s:state bi) =
  let s = deref h s in
  let State block_state buffer total_len seen key = s in
  let blocks, rest = split_at_last (U32.v bi.block_len) seen in
  (* omitted *) ... /\
  bi.state.v h block_state == bi.update_multi_s (bi.init_s (optional_reveal h key)) 0 blocks /\
  slice (as_seq h buffer) 0 (length rest) == rest
\end{minted}

\son{one reviewer was ``lost in this section'' and in particular did not understand why we
need to copy the state. I'm not sure it's worth rephrasing.}
The \li`finish` function takes a block specification \li`bi`.
Under the hood, it calls \li`State.copy` to avoid invalidating
the \li`block_state`; then \li`update_last` followed by \li`finish`, the last two transitions
of \fref{block-machine}. Thanks to the correctness lemmas in the block API along
with the invariant, \li`finish` states that the digest written in \li`dst` is the
result of applying the full block algorithm to the data that was fed into the streaming
state so far.

\begin{minted}{fstar}
[@ Specialize ] val finish (bi:block_index): s:state bi -> dst:buffer U8.t{len dst == bi.out_len} ->
  ST unit (requires fun h0 -> ... (* omitted *))
  (ensures fun h0 s' h1 -> ... /\ as_seq h1 dst == bi.spec_s (get_key h0 s) (get_seen h0 s))
\end{minted}

One point of interest is the usage of a \emph{ghost selector} \li`get_seen`, which in any
heap returns the bytes seen so far. We have found this style the easiest to work with,
as opposed to a previous iteration of our design where the user was required to materialize the
previously-seen bytes as a ghost argument to the stateful functions, such as \li`finish` above.
The previous iteration placed a heavy burden on clients, who were required to perform some
syntactically heavy book-keeping
to thread this argument through function calls; the present style is much more
lightweight.\footnote{The anonymous supplement contains an in-depth explanation of the
respective merits of the three styles we considered, in file \li`Hacl.Streaming.Functor.fsti`.}

This streaming API has one limitation, in that we cannot prove the absence of memory
leaks.  This is a fundamental limitation of using the \li`ST` effect in \lowstar.
However, this can be easily addressed with manual code review or off-the-shelf tools, such
as \li`clang`'s \li`-fsanitize=memory`.

\paragraph{A Note on Properly Compiling the State Type}
An interesting technicality is that the state type, as introduced above,
generates runtime casts due to the Letouzey-style extraction pipeline of \fstar,
and as such, does not compile to C. Casts between values of different types and
sizes are admissible when extracting to OCaml, owing to its universal boxed
value representation (as long as one is willing to use \li+Obj.magic+). But C has
no $\top$ type, meaning that such casts are rejected by \krml.

Looking closely at \li+state_s+, we remark that it is parameterized by a value,
not a type. It therefore won't extract to a definition of the form
\li+type 'a t+. Second, it uses a type-level field projection for
\li+block_state+, which
is also not part of the simple grammar of types of either OCaml or C.

We \emph{do} instantiate \li+state_s+ over a specific choice of argument
\li+bi+. But inductive types are typed nominally, and an application of
\li+state_s+ to its argument generates a type instantiation, not a fresh,
specialized state type definition. This is in contrast to a type abbreviation,
which, being typed structurally, would simply reduce away, circumventing this issue.

We could rewrite this type too, using our tactic, but there is actually a
simpler way. We add a seemingly useless type (\emph{not} value!) parameter to
\li+state_s+:.

\begin{minted}{fstar}
[@ CAbstractStruct] type state_s' (bi: Block.block_index) (s: Type { s == bi.state.s }) =
  { block_state: s; ... (* rest as before *) }

let state_s bi = state_s' bi bi.state.s
\end{minted}

From the point of view of type-checking, this is strictly equivalent to the
previous definition. But from the point of view of extraction, after erasure,
\li+bi+ becomes an unused, erased type parameter of \li+state_s'+ (it eventually
gets eliminated), while \li+s+, at \li+Type+, becomes a regular parameter of
the (extracted) data type \li+state_s'+. Uses of \li+state_s'+, via the
\li+state_s+ wrapper, become regular type applications. This means that the
resulting code contains no casts, and simply relies on parameterized data types,
which are handled by \krml and monomorphized via a whole-program compilation
pass.

This rewriting trick significantly improves the quality of the generated code,
and to the best of our knowledge, had never been documented before.

\paragraph{A Note on Additional Compile-Time Parameters}
In addition to types and lemmas, we also add, within our index, extra parameters that
reduce at compile-time using normal reduction mechanisms. These act as
supplemental ``tweaking knobs'' that control the shape of the produced code. An
example is the block size, which is specific to each algorithm, reduces
using normal partial evaluation mechanisms, and eventually generates
stack-allocated arrays of the correct block size (rather than with a run-time dynamic
check).

Another one of these knobs is the key management policy, which is another choice
the user can tweak when instantiating the streaming API.

\begin{minted}{fstar}
type key_management = | Runtime | Erased

let optional_key (km: key_management) (key: Key.state_index) : Type =
  match km with | Runtime -> key.s | Erased -> Ghost.erased key.t
\end{minted}

The \li`km` parameter of the \li`block` API exists only at compile-time,
not at run-time. All of its uses are partially evaluated away.
It allows the \li`block` algorithm to indicate whether it needs a key. In the streaming
code, every reference to key goes through a wrapper like the one above. After partial evaluation,
the \li`optional_*` wrappers reduce to either a proper key type, or to a ghost value, which
then gets erased to \li`unit`.
This allows, for instance, generating either an \li+init+ function that does
\emph{not} take a key (hash functionality), or an \li+init+ function that
\emph{does} take a key (MAC functionality). Thanks to the various unit-elimination
optimizations of \krml, the former case results in no superfluous fields in the
state type, nor superfluous arguments to the API functions.

\section{Evaluation}
\label{sec:evaluation}

We now evaluate the efficiency of our approach. Recall that our original goal
was to support authoring large-scale, low-level verified software; in this
section, we therefore focus on proof engineering and programmer productivity metrics.
Our case studies involve pre-existing algorithms from the \haclstar project; the
run-time performance of the code is thus that of the underlying cryptographic
algorithms, for which we did not observe noticeable changes in performance after
we updated the code. We therefore leave a crypto-oriented performance discussion to the
original \haclstar paper~\cite{haclxn}.
In total, the modifications we performed had an impact on 30k lines of the C code
generated by compiling the \haclstar library.

\subsection{Core Algorithms: ChaCha20-Poly1305, Curve25519, HPKE}
\label{sec:evaluation:core-algos}

\paragraph{Qualitative Study}
The \haclstar library originally featured ChaCha20-Poly1305 and Curve25519, in
multiple variants, but got by without the use of our code rewriting tactic. For
Curve25519, the
original code was playing build system tricks, and would tweak the include path
to select, say, one implementation of the Curve field over another. Needless to
say, this did not scale. Every tweak to the include path invalidates
intermediary build files, with two consequences: first, the build time rapidly
skyrockets; second, the limitation carries over to verified clients of \haclstar
which in turn need to play the same include path tricks if they want to use such
algorithms.

In the case of ChaCha20-Poly1305, the existing code was in better shape, but not by
much. It relied on a static dispatch style
(not described here), which came with severe limitations. Notably, it
imposed that all variants of the same algorithm be in one C file. This made
regular C and vectorized
implementations appear in the same file; as the vectorized version would
mandate
special
compiler flags (here,  \li+-mavx -mavx2+), the C compiler would happily use AVX2
instructions for the
non-vectorized, regular C version, causing illegal instruction errors later
on~\cite{evercrypt-oops}.

We upgraded both of these algorithms within the \haclstar codebase to use our
code-rewriting tactic, which addressed all of the roadblocks above, and resulted
in significantly improved programmer experience and productivity.

Our techniques also paved the way for the HPKE implementation in \haclstar.
Before our work, \haclstar could not distinguish between a notion of
\emph{algorithm} (e.g., P-256 \emph{vs.} Curve25519) and multiple
\emph{implementations} (e.g., Curve25519-64 \emph{vs.} Curve25519-51) of said
algorithm.
This made a modular and specializable HPKE
impossible to author. Thanks to our framework, the \haclstar authors were able
to express HPKE naturally, modularly and generically, while allowing more than
60 possible choices of algorithms and corresponding implementations, each in
their own file. This simply could not have happened without the principles
exposed in this article.

\paragraph{Quantitative Study}

In the design of elaborator reflection, the user (i.e., the tactic) is allowed
to create ill-typed terms. The API does not statically enforce the creation of
well-typed terms; it simply re-checks user-provided terms before they are
added to the context. This means that the rewritten terms produced by our tactic
need to be re-checked by the \fstar typechecker.

We measure the verification overhead that comes from re-verifying those rewritten
terms. Specifically, \tref{tactic} measures the overhead incurred by
re-checking the tactic-generated definitions, relative to the total verification
time for a given algorithm. In most cases, the overhead is $<
100\%$, because we don't rewrite lemmas and proofs.
We need to investigate why HPKE is an outlier; we suspect the Z3 solver
might be overly sensitive to the shape of the proof obligations it receives;
since we rewrite the call-graph, the resulting proof obligations are slightly
different from the ones generated by the original call-graph.

One might wonder about the impact of our approach on a programmer's productivity.
Indeed, re-verifying the terms has a non-negligible impact on the build time
which, in turn, is extremely important to programmer productivity.
In practice this did not prove to be an issue, because we generally need fast incremental
builds (and in particular, fast type-checking of the code) when working on the \emph{generic}
definitions and their proofs (i.e., the functions implemented in the DSL), or when working
on the clients of the specialized instantiations \emph{after} we run the call-graph rewriting
and verified the result.

\begin{table}[t]
\begin{minipage}[t]{0.30\textwidth}
  \small
  \centering
\begin{tabular}{|l|l|} \hline
\textbf{Algorithm} & \mbox{\textbf{Verif. time}}\\ \hline
Chacha20 & 6s (+27\%) \\ \hline
Poly1305 & 43s (+17\%) \\ \hline
ChaCha-Poly & 19s (+34\%) \\ \hline
Curve25519 & 74s (+88\%) \\ \hline
HPKE & 231s (+132\%) \\ \hline
\end{tabular}
\caption{Cost of verifying the tactic-rewritten call graphs}
\label{table:tactic}
\vspace{-2em}
\end{minipage}
\begin{minipage}[t]{0.66\textwidth}
  \centering
  \small
\begin{tabular}{|l|l|l|l|l|} \hline
& \textbf{\fstar LoC} & \textbf{C LoC} & \textbf{verif.} & \textbf{extract.} \\

\hline

EverCrypt hashing (old) & 848 & 798 & & \\ \hline \hline
\textbf{API and interfaces} & 1667 & 0 & 103.5s & 0 \\ \hline
EverCrypt hashing (new) & 128 & 929 & 20.3s & 8.3s \\ \hline
MD5, SHA1, SHA2 ($\times 4$) & 231 & 1577 & 39.4s & 13.6s \\ \hline
Poly1305 ($\times 3$) & 452 & 998 & 55.7s & 6.4s \\ \hline
Blake2s ($\times 2$), Blake2b ($\times 2$) & 874 & 4761 & 115.1s & 9.2s \\ \hline
\textbf{Total} & 4200 & 8265 & 334s & 47.6s \\ \hline
\end{tabular}
\caption{Quantifying the impact of the streaming API}
\label{table:streaming-effort}
\vspace{-2em}
\end{minipage}
\end{table}

\subsection{Implementation and Usability of our DSL}

Tactics are not part of the trusted computing base (\sref{background}); unlike,
say, MTac2~\cite{kaiser2018mtac2}, Meta-\fstar~\cite{metafstar}
does not allow the user to prove properties about tactics, trading provable
correctness for ease-of-use and programmer productivity.
This begs the question of the reliability of the tactic, since it's not formally
shown to always generate well-typed terms. Debugging took place in two phases.
First, type-checking the implementation of the tactic itself, which was easy, as there were no deep
proof obligations, only ML-like type-checking. We note that our tactic, at 620
lines, (including whitespace and comments) is the third largest Meta-\fstar
program written to date. Second, type-checking the output of the tactic. We did
so by inspecting the generated definitions and type-checking them like regular
terms in the interactive mode, which quickly revealed the source of bugs.
We debugged the tactic on Curve25519, our most complex example; once debugged,
the tactic never generated ill-typed code and was used successfully by other
co-authors.

In the years since we implemented this tactic, it has come to be used in numerous
places in \haclstar and has been the workhorse of many verified algorithms. The
tactic now executes natively, leveraging the \fstar compiler's ability to
dynlink natively-compiled tactics, similar to Coq's \li+native_compute+. The
running-time of the tactic itself is not noticeable.

\subsection{Streaming API}

To evaluate the applicability of the streaming API, we compare lines of code
(LoC) for the \fstar source code and the final C code as a proxy for programmer
effort. While not ideal, this metric has been used by several other
papers~\cite{hacl,haclxn,evercrypt} and
provides a coarse estimate of the proof engineering effort. Our point of
reference is a previous, non-generic streaming API that previously operated atop
the EverCrypt agile hash layer.

\tref{streaming-effort} presents the evaluation. For the old, non-generic
streaming API, the proof-to-code ratio was $1.11$, i.e., we had to write more
than one line of code in \fstar for every line of generated C code.

Capturing the block API and implementing the streaming API uses 1667 lines of \fstar
code. The extra verification effort is quickly amortized across the 14
applications of the streaming API, which each requires a modest amount of
proofs to implement the exact signature of the block API.
Out of those, six have been integrated into the reference implementation of the Python
programming language.
Poly1305 and Blake2
were originally authored without bringing out the functional, fold-like nature
of the algorithms, which led to some glue code and proofs to meet the block
API. Altogether, we obtain a final proof-to-code ratio of $0.51$, which we
interpret to coarsely mean a 2x improvement in programmer productivity. We
expect this number to further decrease, as more applications of the streaming
API follow.

For execution times, we present the verification time of the API itself, and
the verification time of each of the instances, including glue proofs.
Compared to fully verifying Blake2 (7.5 minutes), or Poly1305 (\textasciitilde
14 minutes), the verification cost is modest.
Applying the streaming API to a type class argument incurs no verification
cost, so the extraction column measures the cost of partial evaluation and
extraction to the ML AST, which is negligible.

\section{Related Work}

Automating the generation of low-level code is a common theme among several
software verification projects, and is by no means specific to the \haclstar
universe. We now review several related efforts not discussed earlier in the paper.

A rich overview of the topic of proof engineering can be found in Ringer et al.'s
survey~\cite{qed_at_large}.
We however note that this survey focuses on techniques for verifying large-scale
\emph{proof} developments without efficient, readable extraction being a
concern. Furthermore, it especially focuses on Interactive Theorem Provers, and leaves out
of scope program verifiers based on constraint solvers, which require a different set of
techniques to tame the solver’s complexity.
In this regard, the present work explores a complementary facet of the art of proof engineering.

Fiat Cryptography~\cite{fiat-crypto} relies on a combination of partial evaluation and
certified compilation phases to compile a generic description of a bignum
arithmetic routine to an efficient, low-level imperative language, which is then
output as either C or assembly. In this approach,
the specifications are declarative, and do not impose any choice
of representation. Conversely, in \haclstar, the decision is made by the
programmer, who \emph{manually} refines a high-level mathematical specification into an
implementation that picks word sizes and representations.
While the approach of Fiat Cryptography is
automated, it relies on fine-grained control of the compilation toolchain; for
instance, a key compilation step is bounds inference, which picks integer widths
to be used by the rest of the compilation phases. By contrast,
we do not customize the extraction procedure of \fstar, nor extend \krml with
dedicated phases. Another difference to highlight is that FiatCrypto, to the best of our
knowledge, focuses on the core bignum subset of operations, and offers neither
a high-level Curve25519 API, or other families of algorithms.
In our work, we operate at higher levels up the stack, tackling high-level API
transformers and complete algorithms, ``in the large''.

Jasmin~\cite{jasmin,jasmin2020} is a framework for developing high-speed, verified
cryptographic implementations. Jasmin provides a low-level DSL with features
such as loops or procedures, and has been
successfully used to verify a range of cryptographic algorithms. However it lacks
the higher-level abstraction features provided by our approach to author generic,
specializable implementations.
Jasmin relies on verified compilation using the Coq proof assistant to generate
optimized assembly code semantically equivalent to code verified in
the Jasmin DSL. In contrast, the extraction procedure of \fstar, in line
with several other proof assistants, is trusted;
this problem is orthogonal to our approach, and could be addressed through
advances in verified program extraction~\cite{letouzey2002new,anand2017certicoq}.

Recent work by Pit-Claudel \emph{et.al.}~\cite{FiatToFacade,rupicola} proposes
correct-by-construction pipelines to generate efficient low-level implementations
from non-deterministic functional high-level specifications.
The process is end-to-end
verified since it relies on Bedrock~\cite{chlipala2013bedrock,bedrock2}. In
Pit-Claudel's work, compilation and extraction are framed as a backwards
proof search and synthesis goal. Handling non-determinism has not been done at
scale with \fstar; however, the algorithms we study are fully deterministic.
Pit-Claudel's approach is DSL-centric: the user is expected to augment the
compiler with new
synthesis rules whenever considering a new flavor of specifications. In our
work, we reuse the existing extraction facility of \fstar, which we treat as a
black box. We rely on many whole-program compilation
phases, such as the various compilation schemes for data types, monomorphization
and unused argument elimination; to the best of our knowledge, Pit-Claudel's toolchains
do not support such whole-program transformations that require making non-local
decisions.

Appel~\cite{Appel:2015:VCP:2764452.2701415} verifies the equivalent of our
streaming API applied to the SHA2-256 block algorithm; specifically, the
version of it found within OpenSSL. This work is end-to-end verified, by virtue
of using VST~\cite{vst}. In its current form, the development is geared towards
SHA-256 only, and supports neither higher-order, modular reasoning, nor code generation
``for free'' for multiple algorithms.

Lammich~\cite{lammich2019refinement} uses an approach very similar to
Pit-Claudel \emph{et.al.}, and refines Isabelle/HOL specifications down to
efficient, Imperative/HOL code. The code is then extracted to a functional
programming language, e.g. OCaml or SML, and compiled by an off-the-shelf
compiler. This means that unlike Pit-Claudel, Lammich still relies on a built-in
extraction facility. This is a promising approach, and seems applicable to the
original \haclstar code: it would be worthwhile to try to refine \haclstar
specifications automatically to the \lowstar code. In the case of the various
``functors'' we describe, however, we suspect ``explaining'' how to refine the
specification into the exact API we want would require the same amount of work
as writing the functors directly.

In Haskell, the cryptonite library~\cite{haskellcryptonite}
offers an abstract hash interface using type
classes, along with several instances of this type class. The high-level idea is
the same: unifying various hash algorithms under a single
interface. One natural advantage of our work is that it comes
with proofs, meaning that clients can be verified on top of \haclstar and shown to not
misuse our APIs, before being also extracted to C.
Perhaps more to the point, our approach also has unique constraints: no
matter how efficient functional code may be, we insist on
generating idiomatic C code: adoption by Firefox, Linux and others comes at that
cost. The code produced by our toolchain cannot afford to have
run-time dictionaries or function pointers. We therefore \emph{must} partially
evaluate away the abstractions before C code generation.

Cogent~\cite{cogent} is a purely functional language with linear types that was
used to implement verified file systems~\cite{cogent_file_system}. By restricting the
language's expressiveness, its authors simplify reasoning about Cogent
programs, and allow compiling such programs to efficient C code by means of a self-certifying
compiler. Though Cogent provides high-level features such as
polymorphic, higher-order functions, as well as the possibility of parameterizing code
with abstract definitions, it doesn't provide the equivalent of 0-cost functors like our
approach.
Our approach seems a natural fit to verify applications such as file systems, provided the
effect system we use is adequate. In this regard, it might be interesting to use
our rewriting mechanism with programs written in \steel~\cite{steel_2021}, a separation logic
framework implemented for \fstar and which also supports extraction to C through \krml;
doing so would only require minor modifications of our rewriting procedure.

\section{Conclusion}

Software verification is entering new territory, with proof developments now
routinely topping 100,000 lines of code. And when it comes to verifying
security-critical code, the resulting software artifact has to be not only
verified, but also low-level and fast. For those projects, there is a growing
need of foundational proof development design patterns.

In this paper, we designed, implemented and evaluated a new methodology that
relies on elaborator reflection to add a custom pre-compilation stage. That
early stage interprets user-provided annotations (in effect, a DSL) and rewrites
the code accordingly. We showed that this provides significant gains in terms of
proof engineer productivity, allowing not only existing algorithms in \haclstar
to be rewritten in a form that tames previous complexity, but also allows us to
explore and analyze new algorithms, such as the streaming API.

The benefits of our approach are very concrete: we were able to implement,
verify, and instantiate the streaming API in a modular way, which resulted
in code that was high-quality enough to pass muster with the Python maintainers,
and should be included in the upcoming Python 3.12.

\clearpage
\bibliographystyle{plain}
\bibliography{evercrypt}

\end{document}

%% file: lstfstar.tex
\lstdefinelanguage{fstar}{%
  morekeywords=[1]{type,and,val,fun,let,in,ref,try,if,then,else,match,with,open,module,rec,end,assume,private,when,forall,ghost,assert,function,logic,array,pattern,effect,requires,ensures,decreases,modifies, abstract},
  morekeywords=[2]{WP, Post, Pre, PURE, DIV, STATE, EXN, ALL, TotST, M, GHOST, BAD, GTot, Lemma},
  morekeywords=[3]{},
  morekeywords=[4]{},
  morestring=[b]",
  sensitive=true,%
  numbersep=4pt,
  columns=[l]fullflexible,
  texcl=true,
  mathescape=true,
  identifierstyle={\sffamily},
  keywordstyle=[1]{\sffamily\maybecolor{dkblue}},
  keywordstyle=[2]{\sffamily\maybecolor{dkblue}},
  keywordstyle=[3]{\maybecolor{dkred}},
  keywordstyle=[4]{\rmfamily\itshape},
  rangeprefix=(*---\ ,
  includerangemarker=false,
  stringstyle=\ttfamily,
  showspaces=false,
  morecomment=[n]{(*}{*)},
  commentstyle={\itshape\maybecolor{dkred}},
  literate={->}{$\rightarrow\,$}{1}
           {<-}{$\leftarrow\,$}{1}
           {<>}{$\neq\,$}{1}
	   {delta}{$\delta$}{1}
	   {exists}{$\exists$}{1}
	   {forall}{$\forall$}{1}
      	   {True}{$\top$}{1}
      	   {False}{$\perp$}{1}           
           {fun}{$\lambda$}{1}
           {\\in}{$\in\,$}{1}
           {~}{$\neg$}{1}           
           {tau}{$\tau\,$}{1}
           {STATE0}{$\mathsf{ST}^\prime\,$}{1}
	   {/\\}{$\wedge\;$}{1}
	   {\\/}{$\vee\;$}{1}
           {>=}{$\geq\ $}{2}
           {<=}{$\leq\ $}{2}
	   {<==>}{$\Longleftrightarrow \ $}{3}
	   {==>}{$\Longrightarrow \ $}{3}
           {_}{\textunderscore}{1}
           {--}{-}{1}   
           ,
  breaklines=false}

%% file: tango-colors.tex
\definecolor{butter1}{HTML}{FCE94F}
\definecolor{butter2}{HTML}{EDD400}
\definecolor{butter3}{HTML}{C4A000}
\definecolor{orange1}{HTML}{FCAF3E}
\definecolor{orange2}{HTML}{F57900}
\definecolor{orange3}{HTML}{CE5C00}
\definecolor{chocolate1}{HTML}{E9B96E}
\definecolor{chocolate2}{HTML}{C17D11}
\definecolor{chocolate3}{HTML}{8F5902}
\definecolor{chameleon1}{HTML}{8AE234}
\definecolor{chameleon2}{HTML}{73D216}
\definecolor{chameleon3}{HTML}{4E9A06}
\definecolor{skyblue1}{HTML}{729FCF}
\definecolor{skyblue2}{HTML}{3465A4}
\definecolor{skyblue3}{HTML}{204A87}
\definecolor{plum1}{HTML}{AD7FA8}
\definecolor{plum2}{HTML}{75507B}
\definecolor{plum3}{HTML}{5C3566}
\definecolor{red1}{HTML}{EF2929}
\definecolor{red2}{HTML}{CC0000}
\definecolor{red3}{HTML}{A40000}
\definecolor{aluminium1}{HTML}{EEEEEC}
\definecolor{aluminium2}{HTML}{D3D7CF}
\definecolor{aluminium3}{HTML}{BABDB6}
\definecolor{aluminium4}{HTML}{888A85}
\definecolor{aluminium5}{HTML}{555753}
\definecolor{aluminium6}{HTML}{2E3436}

%% file: paper.bbl
\begin{thebibliography}{10}

\bibitem{firefox-bug}
Crash in hacl chacha20poly1305\_128 aead\_encrypt \& hacl chacha20poly1305\_128
  aead\_decrypt.
\newblock \url{https://bugzilla.mozilla.org/show_bug.cgi?id=1605369#c20}.

\bibitem{sha2}
{Federal Information Processing Standards Publication 180-4: Secure hash
  standard ({SHS})}, 2012.
\newblock NIST.

\bibitem{CVE-2014-0160}
{CVE}-2014-0160.
\newblock Available from MITRE, {CVE-ID} {CVE}-2014-0160., December 2013.

\bibitem{CVE-2017-5715}
{CVE}-2017-5715. systems with microprocessors utilizing speculative execution
  and indirect branch prediction may allow unauthorized disclosure of
  information to an attacker with local user access via a side-channel
  analysis.
\newblock Available from MITRE, {CVE-ID} {CVE}-2017-5715., January 2018.

\bibitem{CVE-2017-5753}
{CVE}-2017-5753. systems with microprocessors utilizing speculative execution
  and branch prediction may allow unauthorized disclosure of information to an
  attacker with local user access via a side-channel analysis.
\newblock Available from MITRE, {CVE-ID} {CVE}-2017-5753., January 2018.

\bibitem{dm4free}
Danel Ahman, C\u{a}t\u{a}lin Hri\c{t}cu, Kenji Maillard, Guido Mart\'inez,
  Gordon Plotkin, Jonathan Protzenko, Aseem Rastogi, and Nikhil Swamy.
\newblock Dijkstra monads for free.
\newblock In {\em ACM Symposium on Principles of Programming Languages (POPL)},
  January 2017.

\bibitem{jasmin}
Jos{\'e}~Bacelar Almeida, Manuel Barbosa, Gilles Barthe, Arthur Blot, Benjamin
  Gr{\'e}goire, Vincent Laporte, Tiago Oliveira, Hugo Pacheco, Benedikt
  Schmidt, and Pierre-Yves Strub.
\newblock Jasmin: High-assurance and high-speed cryptography.
\newblock 2017.

\bibitem{jasmin2020}
Jos{\'e}~Bacelar Almeida, Manuel Barbosa, Gilles Barthe, Benjamin Gr{\'e}goire,
  Adrien Koutsos, Vincent Laporte, Tiago Oliveira, and Pierre-Yves Strub.
\newblock The last mile: High-assurance and high-speed cryptographic
  implementations.
\newblock In {\em 2020 IEEE Symposium on Security and Privacy (SP)}, 2020.

\bibitem{cogent_file_system}
Sidney Amani, Alex Hixon, Zilin Chen, Christine Rizkallah, Peter Chubb, Liam
  O’Connor, Joel Beeren, Yutaka Nagashima, Japheth Lim, Thomas Sewell, Joseph
  Tuong, Gabriele Keller, Toby Murray, Gerwin Klein, and Gernot Heiser.
\newblock {COGENT}: {Verifying} {High}-{Assurance} {File} {System}
  {Implementations}.
\newblock page~14.

\bibitem{anand2017certicoq}
Abhishek Anand, Andrew Appel, Greg Morrisett, Zoe Paraskevopoulou, Randy
  Pollack, Olivier~Savary Belanger, Matthieu Sozeau, and Matthew Weaver.
\newblock Certicoq: A verified compiler for coq.
\newblock In {\em 3rd International Workshop on Coq for Programming Languages
  (CoqPL)}, 2017.

\bibitem{vst}
Andrew~W. Appel.
\newblock Verified software toolchain.
\newblock In {\em Proceedings of the European Conference on Programming
  Languages and Systems (ESOP/ETAPS)}, 2011.

\bibitem{Appel:2015:VCP:2764452.2701415}
Andrew~W. Appel.
\newblock Verification of a cryptographic primitive: {SHA-256}.
\newblock {\em ACM Trans. Program. Lang. Syst.}, April 2015.

\bibitem{blake2-paper}
Jean-Philippe Aumasson, Samuel Neves, Zooko Wilcox-O'Hearn, and Christian
  Winnerlein.
\newblock {BLAKE2}: Simpler, smaller, fast as {MD5}.
\newblock In {\em Applied Cryptography and Network Security}, pages 119--135,
  2013.

\bibitem{hpke}
R.~Barnes and K.~Bhargavan.
\newblock Hybrid public key encryption.
\newblock IRTF Internet-Draft \url{draft-irtf-cfrg-hpke-02}, 2019.

\bibitem{polybug}
David Benjamin.
\newblock poly1305-x86.pl produces incorrect output.
\newblock
  \url{https://mta.openssl.org/pipermail/openssl-dev/2016-March/006161}, 2016.

\bibitem{hmac}
Lennart Beringer, Adam Petcher, Katherine~Q. Ye, and Andrew~W. Appel.
\newblock Verified correctness and security of {OpenSSL HMAC}.
\newblock 2015.

\bibitem{curve25519}
D.~J. Bernstein.
\newblock Curve25519: New {Diffie-Hellman} speed records.
\newblock In {\em Proceedings of the IACR Conference on Practice and Theory of
  Public Key Cryptography (PKC)}, 2006.

\bibitem{poly1305}
Daniel~J. Bernstein.
\newblock The {Poly1305-AES} message-authentication code.
\newblock In {\em Proceedings of Fast Software Encryption}, March 2005.

\bibitem{everest-snapl}
Karthikeyan Bhargavan, Barry Bond, Antoine Delignat-Lavaud, Cédric Fournet,
  Chris Hawblitzel, Catalin Hritcu, Samin Ishtiaq, Markulf Kohlweiss, Rustan
  Leino, Jay Lorch, Kenji Maillard, Jinyang Pang, Bryan Parno, Jonathan
  Protzenko, Tahina Ramananandro, Ashay Rane, Aseem Rastogi, Nikhil Swamy,
  Laure Thompson, Peng Wang, Santiago Zanella-Beguelin, and Jean-Karim
  Zinzindohou\'e.
\newblock Everest: Towards a verified drop-in replacement of {HTTPS}.
\newblock In {\em Proceedings of the Summit on Advances in Programming
  Languages (SNAPL)}, 2017.

\bibitem{springboard}
Cliff~L. Biffle.
\newblock {NaCl/x86} appears to leave return addresses unaligned when returning
  through the springboard.
\newblock \url{https://bugs.chromium.org/p/nativeclient/issues/detail?id=245},
  January 2010.

\bibitem{polybug2}
Hanno B\"ock.
\newblock Wrong results with {Poly1305} functions.
\newblock
  \url{https://mta.openssl.org/pipermail/openssl-dev/2016-March/006413}, 2016.

\bibitem{vale}
Barry Bond, Chris Hawblitzel, Manos Kapritsos, K.~Rustan~M. Leino, Jacob~R.
  Lorch, Bryan Parno, Ashay Rane, Srinath Setty, and Laure Thompson.
\newblock Vale: Verifying high-performance cryptographic assembly code.
\newblock In {\em Proceedings of the USENIX Security Symposium}, August 2017.

\bibitem{idris}
Edwin Brady.
\newblock Idris, a general-purpose dependently typed programming language:
  Design and implementation.
\newblock {\em Journal of functional programming}, 23(5):552--593, 2013.

\bibitem{chlipala2013bedrock}
Adam Chlipala.
\newblock The bedrock structured programming system: Combining generative
  metaprogramming and hoare logic in an extensible program verifier.
\newblock In {\em Proceedings of the 18th ACM SIGPLAN international conference
  on Functional programming}, pages 391--402, 2013.

\bibitem{moura2008z3}
L.~de~Moura and N.~Bj{\o}rner.
\newblock {Z3}: An efficient {SMT} solver.
\newblock 2008.

\bibitem{lean2015}
Leonardo de~Moura, Soonho Kong, Jeremy Avigad, Floris van Doorn, and Jakob von
  Raumer.
\newblock The {Lean} theorem prover.
\newblock In {\em Proc. of the Conference on Automated Deduction (CADE)}, 2015.

\bibitem{jason-found-a-bug}
Jason~A. Donenfeld.
\newblock new 25519 measurements of formally verified implementations.
\newblock \url{http://moderncrypto.org/mail-archive/curves/2018/000972.html},
  February 2018.

\bibitem{sha3}
Morris~J Dworkin.
\newblock Sha-3 standard: Permutation-based hash and extendable-output
  functions.
\newblock Technical report, 2015.

\bibitem{fiat-crypto}
A.~Erbsen, J.~Philipoom, J.~Gross, R.~Sloan, and A.~Chlipala.
\newblock Simple high-level code for cryptographic arithmetic - with proofs,
  without compromises.
\newblock 2019.

\bibitem{bedrock2}
Andres Erbsen, Samuel Gruetter, Joonwon Choi, Clark Wood, and Adam Chlipala.
\newblock Integration verification across software and hardware for a simple
  embedded system.
\newblock In {\em Proceedings of the 42nd {ACM} {SIGPLAN} {International}
  {Conference} on {Programming} {Language} {Design} and {Implementation}},
  pages 604--619, Virtual Canada, June 2021. ACM.

\bibitem{vale-fstar}
Aymeric Fromherz, Nick Giannarakis, Chris Hawblitzel, Bryan Parno, Aseem
  Rastogi, and Nikhil Swamy.
\newblock A verified, efficient embedding of a verifiable assembly language.
\newblock {\em Proc. ACM Program. Lang.}, 3(POPL), January 2019.

\bibitem{steel_2021}
Aymeric Fromherz, Aseem Rastogi, Nikhil Swamy, Sydney Gibson, Guido Martínez,
  Denis Merigoux, and Tahina Ramananandro.
\newblock Steel: proof-oriented programming in a dependently typed concurrent
  separation logic.
\newblock {\em Proceedings of the ACM on Programming Languages}, 5(ICFP):1--30,
  August 2021.

\bibitem{cpp-tricks}
Paul Futz.
\newblock C preprocessor tricks, tips and idioms.
\newblock
  \url{https://github.com/pfultz2/Cloak/wiki/C-Preprocessor-tricks,-tips,-and-idioms},
  June 2015.

\bibitem{Gu:CertiKOS:2015}
Ronghui Gu, J{\'e}r{\'e}mie Koenig, Tahina Ramananandro, Zhong Shao,
  Xiongnan~(Newman) Wu, Shu-Chun Weng, Haozhong Zhang, and Yu~Guo.
\newblock Deep specifications and certified abstraction layers.
\newblock In {\em Proceedings of the ACM Conference on Principles of
  Programming Languages (POPL)}, 2015.

\bibitem{Gu:2016:CEA:3026877.3026928}
Ronghui Gu, Zhong Shao, Hao Chen, Xiongnan Wu, Jieung Kim, Vilhelm Sj\"{o}berg,
  and David Costanzo.
\newblock Certikos: An extensible architecture for building certified
  concurrent os kernels.
\newblock In {\em Proceedings of the 12th USENIX Conference on Operating
  Systems Design and Implementation}, OSDI'16, pages 653--669, Berkeley, CA,
  USA, 2016. USENIX Association.

\bibitem{aes-gcm-bug}
S.~{Gueron} and V.~{Krasnov}.
\newblock The fragility of {AES-GCM} authentication algorithm.
\newblock In {\em Proceedings of the Conference on Information Technology: New
  Generations}, April 2014.

\bibitem{intelaesni}
Shay Gueron.
\newblock {Intel\textsuperscript{\textregistered} Advanced Encryption Standard
  (AES) New Instructions Set}.
\newblock
  \url{https://software.intel.com/sites/default/files/article/165683/aes-wp-2012-09-22-v01.pdf},
  September 2012.

\bibitem{haskellcryptonite}
Haskell-crypto.
\newblock cryptonite.
\newblock \url{https://github.com/haskell-crypto/cryptonite}.

\bibitem{noisestar}
Son Ho, Jonathan Protzenko, Abhishek Bichhawat, and Karthikeyan Bhargavan.
\newblock Noise*: A library of verified high-performance secure channel
  protocol implementations (long version).
\newblock {\em Cryptology ePrint Archive}, 2022.

\bibitem{awswsj}
The Wall~Street Journal.
\newblock Provable security for modern applications.
\newblock
  \url{https://partners.wsj.com/aws/reinventing-with-the-cloud/provable-security-for-modern-applications/}
  Retrieved February 2023.

\bibitem{kaiser2018mtac2}
Jan-Oliver Kaiser, Beta Ziliani, Robbert Krebbers, Yann R{\'e}gis-Gianas, and
  Derek Dreyer.
\newblock Mtac2: typed tactics for backward reasoning in coq.
\newblock {\em Proceedings of the ACM on Programming Languages}, 2(ICFP):1--31,
  2018.

\bibitem{kassios2006dynamic}
Ioannis~T Kassios.
\newblock Dynamic frames: Support for framing, dependencies and sharing without
  restrictions.
\newblock In {\em 14th International Symposium on Formal Methods}, 2006.

\bibitem{klein2009seL4}
Gerwin Klein, Kevin Elphinstone, Gernot Heiser, June Andronick, David Cock,
  Philip Derrin, Dhammika Elkaduwe, Kai Engelhardt, Michael Norrish, Rafal
  Kolanski, Thomas Sewell, Harvey Tuch, and Simon Winwood.
\newblock {seL4}: Formal verification of an {OS} kernel.
\newblock In {\em Proceedings of the ACM Symposium on Operating Systems
  Principles (SOSP)}, 2009.

\bibitem{cakeML}
Ramana Kumar, Magnus~O. Myreen, Michael Norrish, and Scott Owens.
\newblock {CakeML}: a verified implementation of {ML}.
\newblock In {\em Proceedings of the {ACM} Symposium on Principles of
  Programming Languages (POPL)}, January 2014.

\bibitem{lammich2019refinement}
Peter Lammich.
\newblock Refinement to imperative hol.
\newblock {\em Journal of Automated Reasoning}, 62(4):481--503, 2019.

\bibitem{compcert}
Xavier Leroy, Sandrine Blazy, Daniel K\"astner, Bernhard Schommer, Markus
  Pister, and Christian Ferdinand.
\newblock Compcert -- a formally verified optimizing compiler.
\newblock In {\em Embedded Real Time Software and Systems (ERTS)}. SEE, 2016.

\bibitem{letouzey2002new}
Pierre Letouzey.
\newblock A new extraction for coq.
\newblock In {\em International Workshop on Types for Proofs and Programs},
  pages 200--219. Springer, 2002.

\bibitem{macqueen1986using}
David~B MacQueen.
\newblock Using dependent types to express modular structure.
\newblock In {\em Proceedings of the 13th ACM SIGACT-SIGPLAN symposium on
  Principles of programming languages}, pages 277--286, 1986.

\bibitem{fstar-meta}
Guido Mart\'inez, Danel Ahman, Victor Dumitrescu, Nick Giannarakis, Chris
  Hawblitzel, Catalin Hritcu, Monal Narasimhamurthy, Zoe Paraskevopoulou,
  Cl\'ement Pit-Claudel, Jonathan Protzenko, Tahina Ramananandro, Aseem
  Rastogi, and Nikhil Swamy.
\newblock {Meta-F*}: Metaprogramming and tactics in an effectful program
  verifier.
\newblock In {\em European Symposium on Programming (ESOP)}, 2019.

\bibitem{metafstar}
Guido Mart\'inez, Danel Ahman, Victor Dumitrescu, Nick Giannarakis, Chris
  Hawblitzel, Catalin Hritcu, Monal Narasimhamurthy, Zoe Paraskevopoulou,
  Cl\'ement Pit-Claudel, Jonathan Protzenko, Tahina Ramananandro, Aseem
  Rastogi, and Nikhil Swamy.
\newblock {Meta-F*}: Proof automation with {SMT}, tactics, and metaprograms.
\newblock In {\em 28th European Symposium on Programming (ESOP)}, pages 30--59.
  Springer, April 2019.

\bibitem{mathlib2020}
The mathlib Community.
\newblock The lean mathematical library.
\newblock In {\em Proceedings of the 9th ACM SIGPLAN International Conference
  on Certified Programs and Proofs}, CPP 2020, pages 367--381, New York, NY,
  USA, 2020. Association for Computing Machinery.

\bibitem{gcm}
David~A. McGrew and John Viega.
\newblock The security and performance of the {Galois}/counter mode of
  operation.
\newblock In {\em Proceedings of the International Conference on Cryptology in
  India (INDOCRYPT)}, 2004.

\bibitem{ml}
Robin Milner.
\newblock A theory of type polymorphism in programming.
\newblock {\em Journal of computer and system sciences}, 17(3):348--375, 1978.

\bibitem{sha3-bug}
Nicky Mouha.
\newblock Sha-3 buffer overflow.
\newblock \url{https://mouha.be/sha-3-buffer-overflow/}, October 2022.

\bibitem{mouha2018finding}
Nicky Mouha, Mohammad~S Raunak, D~Richard Kuhn, and Raghu Kacker.
\newblock Finding bugs in cryptographic hash function implementations.
\newblock {\em IEEE transactions on reliability}, 67(3):870--884, 2018.

\bibitem{cogent}
Liam O'Connor, Christine Rizkallah, Zilin Chen, Sidney Amani, Japheth Lim,
  Yutaka Nagashima, Thomas Sewell, Alex Hixon, Gabriele Keller, Toby Murray,
  and Gerwin Klein.
\newblock {COGENT}: {Certified} {Compilation} for a {Functional} {Systems}
  {Language}, January 2016.
\newblock arXiv:1601.05520 [cs].

\bibitem{fast-curve25519}
Thomaz Oliveira, Julio L\'opez, H\"useyin H{\i}\c{s}{\i}l, Armando
  Faz-Hern\'andez, and Francisco Rodr\'iguez-Henr\'iquez.
\newblock How to (pre-)compute a ladder: Improving the performance of {X25519}
  and {X448}.
\newblock In {\em Proceedings of Selected Areas in Cryptography (SAC)}, August
  2017.

\bibitem{openssl-poly1305-bug3}
{OpenSSL}.
\newblock Chase overflow bit on {x86} and {ARM} platforms.
\newblock {GitHub} commit dc3c5067cd90f3f2159e5d53c57b92730c687d7e, 2016.

\bibitem{openssl-poly1305-bug1}
{OpenSSL}.
\newblock Don't break carry chains.
\newblock {GitHub} commit 4b8736a22e758c371bc2f8b3534dc0c274acf42c, 2016.

\bibitem{openssl-poly1305-bug2}
{OpenSSL}.
\newblock Don't loose [sic] 59-th bit.
\newblock {GitHub} commit bbe9769ba66ab2512678a87b0d9b266ba970db05, 2016.

\bibitem{rupicola}
Clément Pit-Claudel, Jade Philipoom, Dustin Jamner, Andres Erbsen, and Adam
  Chlipala.
\newblock Relational compilation for performance-critical applications:
  extensible proof-producing translation of functional models into low-level
  code.
\newblock In {\em Proceedings of the 43rd {ACM} {SIGPLAN} {International}
  {Conference} on {Programming} {Language} {Design} and {Implementation}},
  pages 918--933, San Diego CA USA, June 2022. ACM.

\bibitem{FiatToFacade}
Clément Pit-Claudel, Peng Wang, Benjamin Delaware, Jason Gross, and Adam
  Chlipala.
\newblock Extensible extraction of efficient imperative programs with foreign
  functions, manually managed memory, and proofs.
\newblock In Nicolas Peltier and Viorica Sofronie-Stokkermans, editors, {\em
  Automated Reasoning: 10th International Joint Conference, IJCAR 2020, Paris,
  France, July 1–4, 2020, Proceedings, Part II}, volume 12167 of {\em Lecture
  Notes in Computer Science}, pages 119--137. Springer International
  Publishing, July 2020.

\bibitem{haclxn}
Marina Polubelova, Karthikeyan Bhargavan, Jonathan Protzenko, Benjamin
  Beurdouche, Aymeric Fromherz, Natalia Kulatova, and Santiago
  Zanella-Béguelin.
\newblock Hacl×n: Verified generic simd crypto (for all your favorite
  platforms).
\newblock Cryptology ePrint Archive, Report 2020/572, 2020.
\newblock \url{https://eprint.iacr.org/2020/572}.

\bibitem{evercrypt}
Jonathan Protzenko, Bryan Parno, Aymeric Fromherz, Chris Hawblitzel, Marina
  Polubelova, Karthikeyan Bhargavan, Benjamin Beurdouche, Joonwon Choi, Antoine
  Delignat-Lavaud, C{\'e}dric Fournet, et~al.
\newblock Evercrypt: A fast, verified, cross-platform cryptographic provider.
\newblock In {\em 2020 IEEE Symposium on Security and Privacy (SP)}, pages
  634--653, 2019.

\bibitem{lowstar}
Jonathan Protzenko, Jean-Karim Zinzindohou\'e, Aseem Rastogi, Tahina
  Ramananandro, Peng Wang, Santiago {Zanella-B\'eguelin}, Antoine
  Delignat-Lavaud, Catalin Hritcu, Karthikeyan Bhargavan, C\'edric Fournet, and
  Nikhil Swamy.
\newblock Verified low-level programming embedded in {F*}.
\newblock {\em {PACMPL}}, ({ICFP}), September 2017.

\bibitem{everparse}
Tahina Ramananandro, Antoine Delignat-Lavaud, C{\'e}dric Fournet, Nikhil Swamy,
  Tej Chajed, Nadim Kobeissi, and Jonathan Protzenko.
\newblock Everparse: verified secure zero-copy parsers for authenticated
  message formats.
\newblock In {\em {USENIX} Security Symposium}, pages 1465--1482, 2019.

\bibitem{indexedeffects}
Aseem Rastogi, Guido Mart\'inez, Aymeric Fromherz, Tahina Ramananandro, and
  Nikhil Swamy.
\newblock Programming and proving with indexed effects, July 2021.

\bibitem{qed_at_large}
Talia Ringer, Karl Palmskog, Ilya Sergey, Milos Gligoric, and Zachary Tatlock.
\newblock {QED} at {Large}: {A} {Survey} of {Engineering} of {Formally}
  {Verified} {Software}.
\newblock {\em Foundations and Trends® in Programming Languages},
  5(2-3):102--281, 2019.
\newblock arXiv:2003.06458 [cs].

\bibitem{rossberg2014f}
Andreas Rossberg, Claudio Russo, and Derek Dreyer.
\newblock F-ing modules.
\newblock {\em Journal of functional programming}, 24(5):529--607, 2014.

\bibitem{blake2}
M-J. Saarinen and J-P. Aumasson.
\newblock The blake2 cryptographic hash and message authentication code (mac).
\newblock IETF RFC 7693, 2015.

\bibitem{sewell2013translation}
Thomas Arthur~Leck Sewell, Magnus~O. Myreen, and Gerwin Klein.
\newblock Translation validation for a verified os kernel.
\newblock In {\em Proceedings of ACM PLDI}, 2013.

\bibitem{mumon}
Nikhil Swamy, C\u{a}t\u{a}lin Hri\c{t}cu, Chantal Keller, Aseem Rastogi,
  Antoine Delignat-Lavaud, Simon Forest, Karthikeyan Bhargavan, C\'{e}dric
  Fournet, Pierre-Yves Strub, Markulf Kohlweiss, Jean-Karim Zinzindohou\'e, and
  Santiago {Zanella-B\'eguelin}.
\newblock Dependent types and multi-monadic effects in {F*}.
\newblock In {\em Proceedings of the ACM Conference on Principles of
  Programming Languages (POPL)}, 2016.

\bibitem{everparse3d}
Nikhil Swamy, Tahina Ramananandro, Aseem Rastogi, Irina Spiridonova, Haobin Ni,
  Dmitry Malloy, Juan Vazquez, Michael Tang, Omar Cardona, and Arti Gupta.
\newblock Hardening attack surfaces with formally proven binary format parsers.
\newblock In {\em Proceedings of the 43rd ACM SIGPLAN International Conference
  on Programming Language Design and Implementation (PLDI '22)}, June 2022.

\bibitem{bitcoin-merkle-bug}
Forrest Voight.
\newblock {CVE-2012-2459} (block merkle calculation exploit).
\newblock \url{https://bitcointalk.org/?topic=102395}, August 2012.

\bibitem{evercrypt-oops}
Guido Vranken.
\newblock Cryptofuzz.
\newblock \url{https://bugs.chromium.org/p/oss-fuzz/issues/detail?id=14822#c3},
  May 2019.

\bibitem{warren2013hacker}
Henry~S Warren.
\newblock {\em Hacker's delight}.
\newblock Pearson Education, 2013.

\bibitem{weeks2006whole}
Stephen Weeks.
\newblock Whole-program compilation in mlton.
\newblock {\em ML}, 6:1--1, 2006.

\bibitem{hacl}
Jean~Karim Zinzindohou\'{e}, Karthikeyan Bhargavan, Jonathan Protzenko, and
  Benjamin Beurdouche.
\newblock {HACL*}: A verified modern cryptographic library.
\newblock 2017.

\end{thebibliography}
